\documentclass[journal, twocolumn, 10pt]{IEEEtran}

\usepackage[T1]{fontenc}
\usepackage{lipsum}
\usepackage{blindtext}
\usepackage{multicol}
\usepackage{graphicx}
\usepackage{cite}
\usepackage{amssymb,amsmath}
\usepackage{algorithm}
\usepackage{algpseudocode}
\usepackage{flushend}
\usepackage{multirow}
\usepackage{hhline}
\usepackage{tcolorbox}
\usepackage{graphicx}
\usepackage{cite}
\usepackage{amssymb,amsmath}
\usepackage{algorithm}
\usepackage{color,soul,colortbl}
\usepackage{cleveref}
\usepackage{subcaption}
\usepackage{booktabs}
\usepackage{xcolor}
\usepackage{caption}
\usepackage{multirow}
\usepackage{hhline}
\usepackage{psfrag}

\usepackage{widetext}
\usepackage{flushend}
\usepackage{cuted}
\usepackage{float}
\usepackage{lipsum}

\newcounter{tempEquationCounter}
\newcounter{thisEquationNumber}

\usepackage{amsthm}
\newtheorem{theorem}{Theorem}

\theoremstyle{plain}

\theoremstyle{plain}

\theoremstyle{plain}

\providecommand{\lemmaname}{Lemma}
\providecommand{\propositionname}{Proposition}
\providecommand{\theoremname}{Theorem}
\providecommand{\lemmaname}{Lemma}
\providecommand{\propositionname}{Proposition}
\providecommand{\theoremname}{Theorem}
\makeatother
\providecommand{\lemmaname}{Lemma}
\providecommand{\propositionname}{Proposition}
\providecommand{\theoremname}{Theorem}

\definecolor{G}{gray}{0.9}
\definecolor{LC}{rgb}{0.88,1,1}

\usepackage[nolist]{acronym}
\begin{document}
\title{{\LARGE{An Interdisciplinary Approach to Optimal Communication and Flight Operation of High-Altitude Long-Endurance Platforms}}}

\author{{ Sidrah Javed,~\IEEEmembership{Student~Member,~IEEE,}  Mohamed-Slim Alouini,~\IEEEmembership{Fellow,~IEEE}, and Zhiguo Ding,~\IEEEmembership{Fellow,~IEEE}} 
 \thanks{S. Javed and M.-S. Alouini are with CEMSE Division,  King Abdullah University of Science and Technology (KAUST),  Thuwal, Makkah Province, Saudi Arabia.   E-mail: \{sidrah.javed, slim.alouini\}@kaust.edu.sa }
 \thanks{Zhiguo Ding is with Department of Electrical and Electronic Engineering, The University of Manchester, UK. E-mail: zhiguo.ding@manchester.ac.uk } }
 \maketitle
 \begin{acronym}
\acro{5G}{fifth generation}
\acro{BER}{bit error rate}
\acro{AWGN}{additive white Gaussian noise}
\acro{CDF}{cumulative distribution function}
\acro{CSI}{channel state information}
\acro{FDR}{full-duplex relaying}
\acro{HDR}{half-duplex relaying}
\acro{IC}{interference channel}
\acro{IGS}{improper Gaussian signaling}
\acro{MHDF}{multi-hop decode-and-forward}
\acro{SIMO}{single-input multiple-output}
\acro{MIMO}{multiple-input multiple-output}
\acro{MISO}{multiple-input single-output}
\acro{MRC}{maximum ratio combining}
\acro{PDF}{probability density function}
\acro{PGS}{proper Gaussian signaling}
\acro{RSI}{residual self-interference}
\acro{RV}{random vector}
\acro{r.v.}{random variable}
\acro{HWD}{hardware distortion}
\acro{cHWD}[HWD]{Hardware distortion}
\acro{AS}{asymmetric signaling}
\acro{GS}{geometric shaping}
\acro{PS}{probabilistic shaping}
\acro{HS}{hybrid shaping}
\acro{SISO}{single-input single-output}
\acro{QAM}{quadrature amplitude modulation}
\acro{PAM}{pulse amplitude modulation}
\acro{PSK}{phase shift keying}
\acro{DoF}{degrees of freedom}
\acro{ML}{maximum likelihood}
\acro{MAP}{maximum a posterior}
\acro{SNR}{signal-to-noise ratio}
\acro{SCP}{successive convex programming}
\acro{RF}{radio frequency}
\acro{CEMSE}{Computer, Electrical, and Mathematical Sciences and Engineering}
\acro{KAUST}{King Abdullah University of Science and Technology}
\acro{DM}{distribution matching}
\end{acronym}
\vspace{-5cm}
\begin{abstract}
Aerial communication platforms, stratospheric high-altitude platform stations (HAPS), have the potential to provide/catalyze advanced mobile wireless communication services with its ubiquitous connectivity and ultra-wide coverage radius. Recently, HAPS has gained immense popularity - achieved primarily through self-sufficient energy systems - to render long-endurance characteristics. The photo-voltaic cells mounted on the aircraft harvest solar energy during the day, which can be partially used for communication and station-keeping, whereas, the excess is stored in the rechargeable batteries for the night time sustenance. We carry out an adroit power budgeting to ascertain if the available solar power can simultaneously and efficiently self-sustain the requisite propulsion and communication power expense. We further propose an energy optimum trajectory for station-keeping flight and non-orthogonal multiple access (NOMA) for  multi-cell users, which are served by the directional beams from HAPS communication systems. We design optimal power allocation for downlink (DL) NOMA users along with the ideal position and speed of flight with the aim to maximize sum data rate during the day and minimize power expenditure during the night, while ensuring quality of service. Our findings reveal the significance of joint design of communication and aerodynamics parameters for optimum energy utilization and resource allocation. 
\end{abstract}
\section{Introduction} 

A new paradigm of wireless communication, the sixth-generation (6G) system, envisions an all-coverage network capable of providing ubiquitous connections for space, air, ground and underwater. Aerial communications have the potential to address numerous limitations of terrestrial and satellite communications in order to meet the ever-increasing communication demands. An aerial network is superior to terrestrial infrastructure with features, such as flexible deployment, short-range line-of-sight links, larger coverage area,  resilience to certain natural disasters, and  many degrees of freedom with controlled mobility \cite{tariq2020speculative}. Moreover, HAPS enjoy the advantages like favorable channel conditions, almost stationary positions, smaller footprint for higher area throughput, long-endurance, reduced round-trip delay, and easy maintenance/reusability in relation to satellites. These inherent characteristics enable aerial communication systems to serve unserved or underserved areas by assisting the existing ground and space infrastructure. For instance, they can complement the existing terrestrial networks in bridging the digital divide by  
connecting the unconnected in less privileged, sparsely populated regions or difficult terrains. These floating base stations are particular useful for effective disaster management and rapid relief operation. Moreover, air-based mobile coverage solutions are remedial for dense metropolitan areas, when used
as a relay and/or back-haul for the prevailing infrastructures \cite{alsharoa2019facilitating}. Similarly, they can act as relays between satellites, substituting ground stations, and distributed data-centers for recording the orbital path of satellites. 

High-Altitude Platform Station (HAPS) has emerged as a viable, promising and versatile candidate for aerial communication owing to its unmanned and  disaster-resilient operation, long-endurance characteristics, self-reliance capability, and ultra-wide coverage area.  The key features of the HAPS include seamless merger with the existing communication infrastructure, no special requirements on the operational spectrum/user equipment (UE), ability to adjust and prioritize capacity or coverage, rapid deployment, flexible maintenance/up-gradation, and scaling capabilities \cite{shibata2019study}. 
To summarize, HAPS is a state of the art preference for ubiquitous connectivity and ultra-wide coverage with low-delay characteristics, higher throughput, and minimal power requirement \cite{karapantazis2005broadband}. 
HAPS can provide universal coverage with dedicated/shared broadband connectivity offering numerous use cases including
greenfield coverage, internet of things (IoT), fixed wireless access/catastrophe management, public/private connectivity, and terrestrial backhaul \cite{GSMA-HAPS-2021}.

The concept of HAPS has been developed for last three decades; however, the recent advances in communication technologies, solar panel efficiency, lightweight composite materials, autonomous avionics and antenna beamforming are the prime enablers behind HAPS
\cite{kurt2021vision,anjum2015terahertz}. They are parts of the 3GPP defined non-terrestrial network (NTN) infrastructure for 5G networks and beyond \cite{anicho2021multi}. A solar powered HAPS is capable of cruising in a station-keeping trajectory for several months because of the favorable conditions in lower stratosphere (typically $18$-$24$km above the sea level) i.e., high solar radiance, minimal weather disturbance, mild turbulence and low wind speed \cite{karapantazis2005broadband}. 
HAPS can be realized as a balloon, airship, or aircraft, according to the required power, use cases and cargo capabilities \cite{d2016high}. Although balloons are small, light weight, least complex and cost-effective options, they cannot stay stationary over a specific area. On the other hand, complex fixed wings and airship designs can maintain fixed positions with large cargo and power capabilities ensuring long-endurance.

The key technological challenges in the HAPS implementation are spread across the domains of aerodynamics, energy and communications. Aerodynamics design should have the characteristics like the lightweight structure, durable and weather resistant model, desired payload capabilities, precise positioning, optimal altitude, paradigmatic station-keeping flight trajectory, and reliable unmanned navigation systems for long-endurance etc \cite{GSMA-HAPS-2021}. 
Likewise, self sustaining HAPS requires proficient solar energy harvesting models for maximum irradiance collection, efficient conversion, battery storage and delivery as per the propulsion and communication power requirements. Numerous sensing/monitoring and communication applications demand respective payload carriage capacities for mounted sensors and base stations. Immaculate HAPS operation requires precise power budget to serve as a relay, backhaul and/or a distributed data analytic center. Communication frameworks need impeccable link budgeting, sophisticated multiple access techniques, network topology and interference management with beamforming to serve the users in its broad coverage area. This work is an effort to harmonize and address the aforementioned challenges to meet the expected communication demands. 

 Numerous contributions have been presented to tackle different challenges of HAPS communication systems, such as trajectory optimization, resource management, performance evaluation, and  link/power budgeting \cite{FBmarriott2020trajectory,azzahra2019noma,ji2020energy,
Nokiahsieh2020uav,nauman2017system}.
A recent study has suggested optimal flight trajectory of a solar powered high-altitude long-endurance aircraft encompassing ascend and descend to maximize sunlight exposure and minimize power expenditure, respectively, according to the time of the day \cite{FBmarriott2020trajectory}. Another contribution proposed non-orthogonal multiple access (NOMA) for effective resource management over  millimeter-wave frequency \cite{azzahra2019noma}.
Some other study employed a steerable adaptive antenna array to continuously serve same cell-users during the repetitive flight pattern over its coverage area \cite{Nokiahsieh2020uav}. Moreover, link budgeting with QPSK for varying code rates was carried out to minimize error rates and power consumption \cite{nauman2017system}. However, to the best of authors' knowledge, there is no contribution for jointly addressing the flight operation and communication aspects in a HAPS communication system. 

%
We present an interdisciplinary approach involving atmospheric science, aerodynamics, wireless communications, and electrical energy to mutually tackle the previously mentioned challenges for optimal flight and communication operations with effective resource allocation for long-endurance.

\begin{itemize}
\item We employ solar position algorithms to quantify the available solar energy for a stratospheric HAPS flying on a certain Julian Day at a given time and location. This enables the calculation of harvested electrical energy after incorporating conversion and storage efficiency of the participating solar panels and storage batteries, respectively.  
\item We utilize rigorous mathematical models to estimate the power expenditure of HAPS. The propulsion and accessory consumption depends upon the HAPS dimensions, trajectory,
stratospheric characteristics, and avionics. Similarly, the communication payload power is determined according to the employed transmission equipment, coverage area, service type, access schemes and the desired link targets.  
\item We propose power and spectrum efficient NOMA based multi-cell communications along with the polygonal beam-steering antenna array \cite{Nokiahsieh2020uav} to tackle incurred small-scale/large-scale fading and propagation losses.
\item The day time allows maximum sun exposure enabling maximum available transmission power to serve numerous cellular/IoT users. On the other side, transmission and flight with minimal energy requirements are the preferred choice during night.  Hence, we jointly optimize the flight and communication parameters to maximize sum rate and minimize power consumption while guarantying quality-of-service (QoS).
\item We develop two different iterative algorithms for day and night time operations delivering the closed-form solutions of altitude, airspeed and power factors.
\item The HAPS continues to cruise in the circular trajectory at the derived altitude with the optimal speed for a given time duration and then transitions to the next state while providing the DL-NOMA service link to the users with optimized power allocation. The cycle repeats everyday to maintain net energy balance greater than zero ensuring ubiquitous connectivity, self-sustenance and long-endurance.
\end{itemize}

 \begin{figure}[t]
    \centering
   \includegraphics[width=0.9\linewidth]{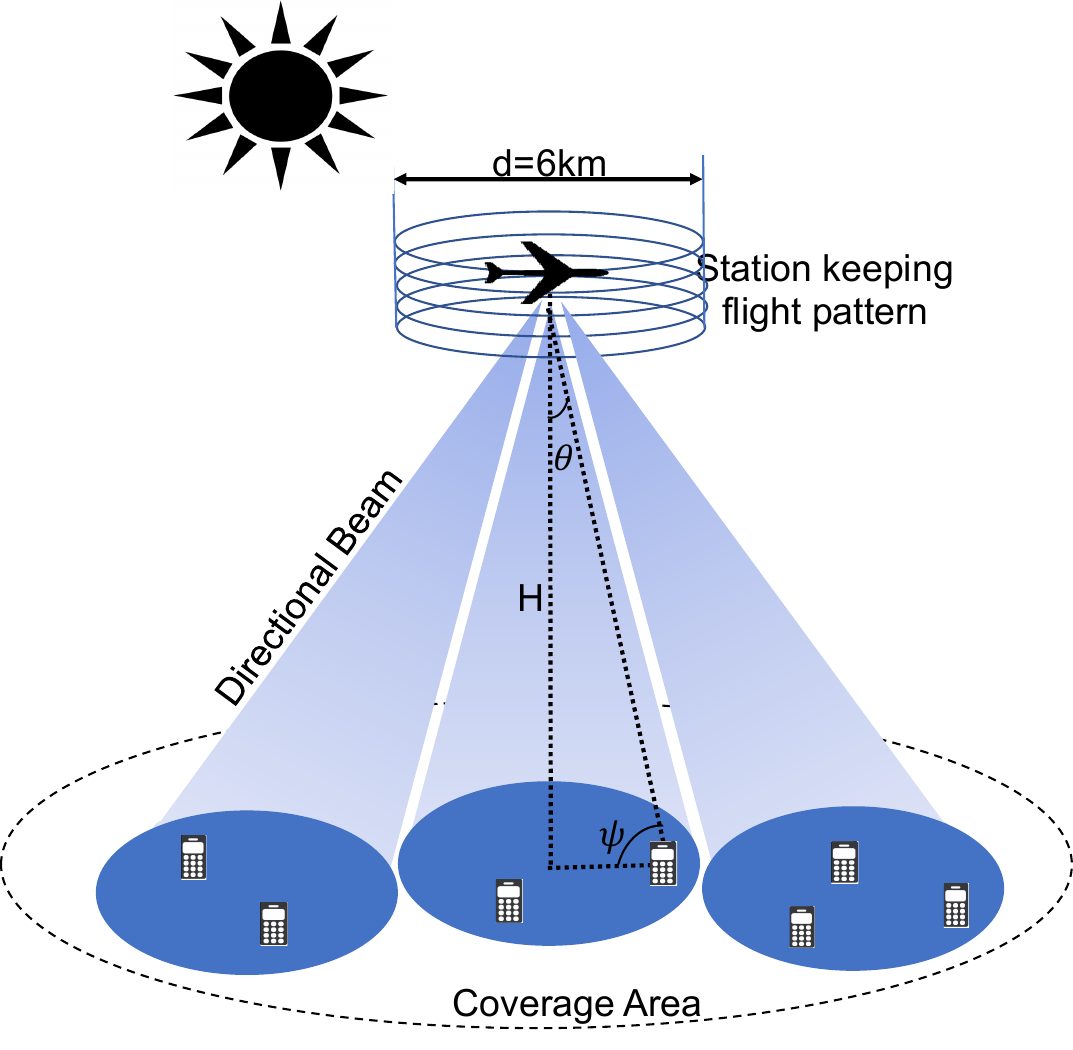}
    \caption{ UAV-based HAPS Multicell Communication System  }
    \label{fig:HAPS}
\end{figure}

The rest of the paper is organized as: Section II describes different aspects of the system under consideration i.e., solar irradiance model, aerodynamics, communication payload and energy storage. 
In section III, we present the sum rate analysis of the users in service area based upon the adopted propagation model and link budgeting.  Next, we formulate the optimization problems to maximize system throughput and minimize power consumption during the day and night time operations, respectively. We present the alternate optimization algorithms to jointly tackle the communication and flight operation in section IV. Later, an illustrative numerical analysis is carried out in section V, to quantify the gains obtained with the proposed scheme, followed by the comprehensive conclusion in Section~VI. 

 In this paper, $|a|$ and $a^*$ represent the absolute and complex conjugate of a scalar complex number $a$. The scalar valued $F(x,y)$ is a function of the independent input variables $x$ and $y$, whereas $F^*$ demonstrates the optimized value of the respective function. The exponential and logarithmic function of variable $z$ are denoted by $\exp(z)$ and $\log(z)$. The summation of different variables or functions over the index $k$ is illustration by  the $\Sigma_k$ sign. 
 A set is presented as $m \in \{ 1,2,\ldots,M\}$ where $m$ can take any value ranging from $1$ to $M$. Moreover, the notation $\{m \setminus n\}$ indicates any value of $m$ in the described range less $n$. 
  On the other hand, a piece-wise function is represented as 
$G(s) =
\begin{cases} 
G_1(s),  & A \\
G_2(s), & B \\
   \end{cases}$
   ,where function $G(s)$ can take the form of $G_1(s)$ when condition $A$ is satisfied or $G_2(s)$ if condition $B$ is true. 
 Furthermore, $x^{(k)}$ and $\mathbf{p}^{(k)}$ represent the instance values of the  variable $x$ and  vector $\mathbf{p}$, respectively, in the $k^{\rm th}$ iteration of an algorithm.
 The notation $Y[n]$ is the value of function $Y$ at a time instance $n$.   The terminology $\frac{\partial P }{\partial v}$ describes the partial derivative of function $P$ with respect to the scalar variable $v$.


\section{System Description}\label{SecII}
We consider a typical unmanned HAPS which follows a station-keeping circular trajectory of diameter $d$ with true airspeed $V$ at an altitude $H$ for station-keeping over the desired service area as shown in Fig. \ref{fig:HAPS}. The stratospheric location (between 18km to 24km) of the HAPS is preferable because of lower wind speed and suitable air density assisting a stable flight operation. In addition, it enables an ultra-wide coverage area with radius from 60 km to 400 km. HAPS provides communication services to various users over the 4G LTE or 5G NR air interface via service link and backhaul to the gateway through the feeder link \cite{Nokiahsieh2020uav}. 
Solar HAPS are appealing owing to their long-endurance characteristics, which can only be achieved if they can harvest more
energy than they expend. Solar cells are mounted on the wings and stabilizers to harvest the abundant solar energy during the day for its routine operation and on-board fuel cells are utilized to store the excess energy for night time sustenance. This section describes the solar irradiance, energy storage, aerodynamics and communication  models with the aim to quantify the availability, storage and consumption of electrical power. The objective is to achieve ubiquitous connectivity with minimal power consumption and enhanced system performance.

%
 
%
%
%

\subsection{Solar Irradiance Model}
Long-endurance platform solely depends on the harvested solar power for its flight and communication operations. Thus, it is imperative to adopt an accurate solar irradiance model to quantify the solar flux
at the surface of the mounted photo-voltaic cells. 
Solar flux is a measure of light energy that is being radiated at a certain area, given in $W/m^2$. The amount of solar flux depends on the Julian day index ($j_d$) and solar elevation angle ($\epsilon_s$) at a given time and location \cite{reda2004solar}.  The extra-terrestrial radiations undergo attenuation (due to Rayleigh scattering and molecular absorption) while traveling through the atmosphere before falling on the surface. However, the solar radiation intensity is still higher in the stratosphere relative to the Earth's surface because of less gaseous and water vapour absorption. Interestingly, stratosphere bears only $0.05\%$ of the amount of water vapor on the Earth’s surface \cite{brizon2015solar}. The solar irradiance at an altitude $H$ is adjusted to account for annual variation due to eccentricity of Earth's orbit and atmospheric absorption factor $f\left(H, \epsilon_s\right)$ as \cite{FBbolandhemmat2019energy}
\begin{equation}\label{eqSI}
I(H,dt) = I_0 \left(  1+0.034 \cos \frac{2\pi j_d (dt)}{365}   \right) f\left(H, \epsilon_s (dt) \right),
\end{equation}

where $I_0$ is a standard solar constant at zero air mass with value $1367$ W/$\rm{m}^2$ defined by American Society Testing and Materials ASTM E490, $j_d$ is the Julian day and $f\left(H, \epsilon_s\right)$ is given as 
\begin{equation}
f\left(H, \epsilon_s (dt)\right) = \exp \left( -p_{\rm R} (H) {\rm m}_{\rm R} (90^0 - \epsilon_s (dt)) \alpha_{\rm ext} (dt) \right),
\end{equation}
\begin{figure}[t]
    \centering
   \includegraphics[width=\linewidth]{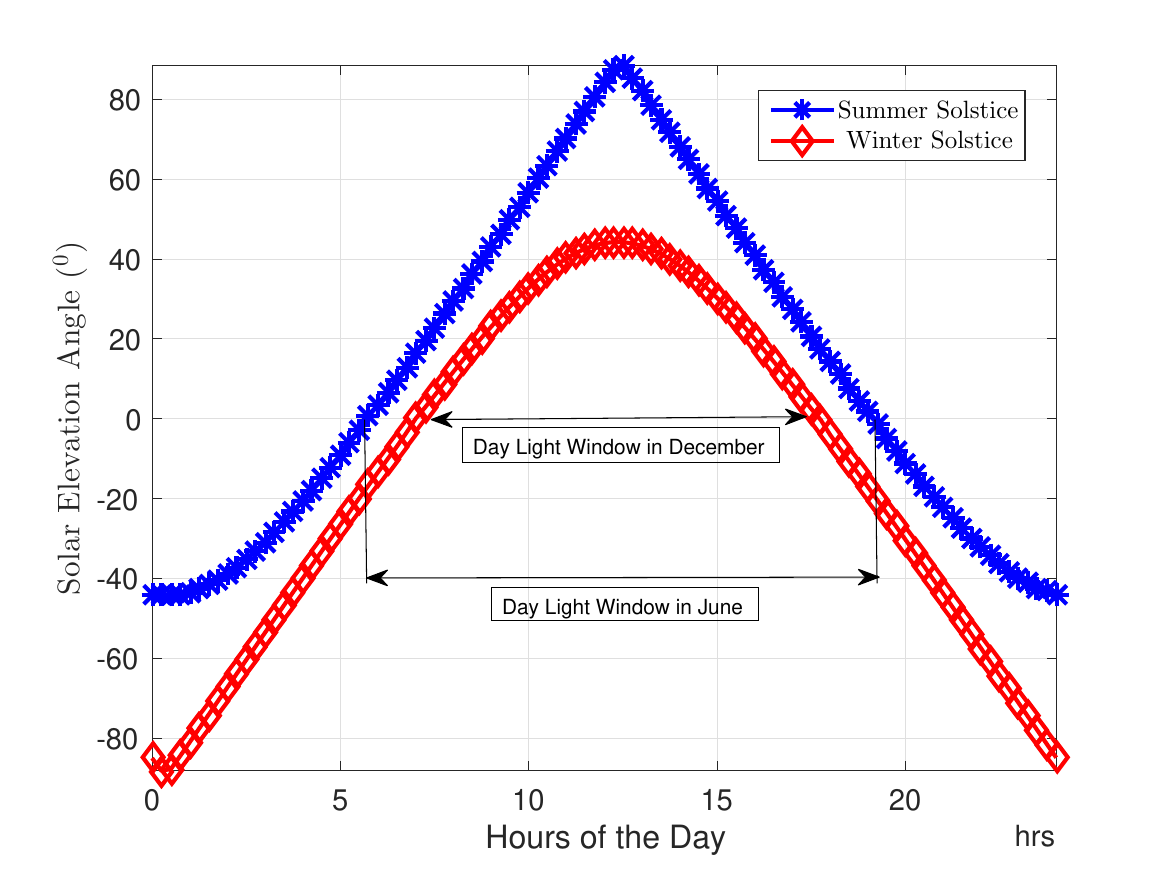}
    \caption{Solar elevation angle variation with hours of the day. }
    \label{fig:SEA}
\end{figure}

where the extinction parameter for clear atmosphere $\alpha_{\rm ext} (dt)$  and relative air mass ${\rm m}_{\rm R}(\theta)$ are taken from \cite{aglietti2009harnessing}. The solar elevation angle $\epsilon_s (dt)$ is computed for a given longitude $\xi$, latitude $\chi$ and date/time $dt$ using the solar position algorithm \cite{reda2004solar}. For this purpose, $\xi$ and $\chi$ are computed from north-position, east-position, and altitude of the HAPS using the WGS84 Earth model. For instance, Fig. \ref{fig:SEA} demonstrates the solar elevation angles throughout the day on summer solstice (SS) and winter solstice (WS) in the year 2023 over the WGS84 location coordinates ($22.3095^{\circ}$,$39.1047^{\circ}$). The positive elevation angles represent position of the sun above the horizon and vice versa. Solar elevation angle is the maximum at noon as anticipated, however, the difference of 
$44.2^{\circ}$ between maximum elevations $\epsilon_{sm}$ of SS and WS is particularly significant for self-sustaining operation throughout the year. The relative pressure $p_{\rm R} (H)=\frac{p_{h}}{p_0}$ at an altitude $H$ is calculated according to the international standard atmosphere (ISA) and 1976 U.S. Standard Atmosphere (USSA) \cite{united1976us}

\begin{figure}[t]
    \centering
   \includegraphics[width=\linewidth]{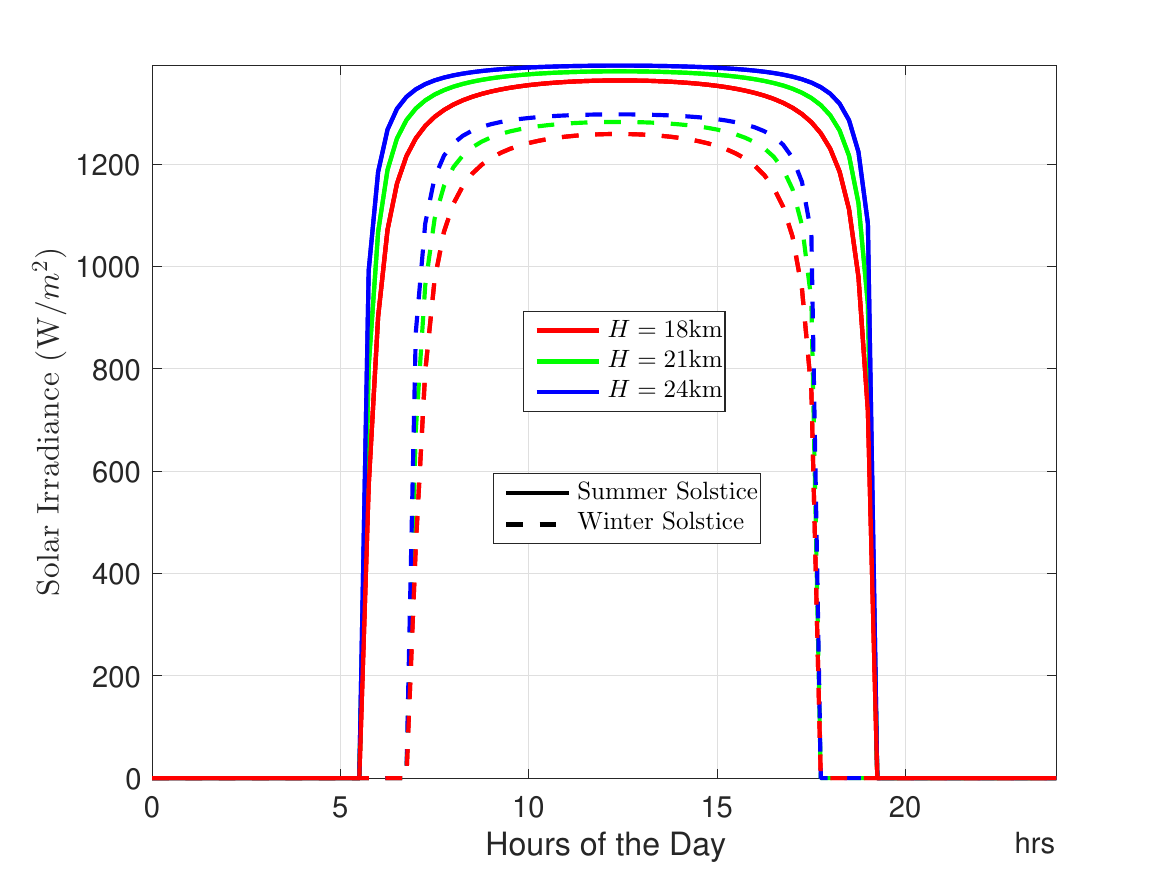}
    \caption{Solar flux per unit area versus hours of the day for three different flight altitudes. }
    \label{fig:SI}
\end{figure}

\begin{equation}\label{eq.rel_pressure}
p_{\rm h} \left(H\right) =
\begin{cases} 
{p_{b1}} \exp \left( \frac{-g M \left(  H-H_{b1} \right)}{RT_{b}}     \right),  & 11km \leq H \leq 20km \\
{p_{b2}}  \left(    \frac{T_b}{T_b+L_b(H-H_{b2})} \right)^{\frac{g M}{RL_b}},  & 20km \leq H \leq 32km \\
   \end{cases}
\end{equation}
where pressure at mean sea level $p_{0}$ is $101325 {\rm Pa}$, base altitudes are $H_{b1}=11 {\rm km} $ and $H_{b2}=20 {\rm km} $ with base static pressures
$p_{b1}=22632.06 {\rm Pa} $ and $p_{b2}= 5474.889 {\rm Pa}$, respectively. Base temperature $T_{b}$ is $216.65 {\rm K}$ with base temperature lapse rate $L_b=1 {\rm K/km}$, 
universal gas constant is $R = 8.31432 {\rm N.m/mol.K}$,
gravitational acceleration $g$ is $9.8{\rm m}/{\rm s}^2$, and molar mass of Earth's air $M$ is taken as $0.0289644 {\rm kg/mol}$.

The average available solar irradiance at each hour of the day on SS and WS of the year $2023$ at location ($\chi$, $\xi$) are presented in Fig. \ref{fig:SI}. Intuitively, the solar flux is only available during the day hours. It increases with the increasing $\epsilon_s$ reaching its maximum value around noon and then decreases with the decreasing $\epsilon_s$. The window of daylight hours is relatively smaller for the winter months i.e., $6.75$hrs-$17.75$hrs as opposed to the summer months i.e.,$5.5$hrs-$19.25$hrs rendering a difference of $2.75$hrs. Moreover,  
the figure demonstrates lower solar irradiance levels during the WS as compared to those during SS with $7.34\%$ percentage reduction at $12.5$hrs due to the increased extinction factor and decreased elevation angles.  We further study the variation of solar irradiance levels with the increasing altitude. We observe a relative decrease of $2.12\%$ in solar irradiance with the altitude changing from $24$km to $18$km. Hence, the night time operation (NTO) is limited by the stored power, whereas the day time operation (DTO) can utilize all the surplus power for its payload. Interestingly, the extended daylight hours in the summer render a slightly higher power budget for nights as they allow extra energy storage before saturating the batteries.

The radiation intensity varies with the varying solar elevation angle throughout the day, while the total solar radiance is dependent on the number of day light hours  between sunrise and sunset. According to Lambert’s cosine law, the instantaneous irradiance $I_i(H,dt)$ measured on a plane will vary with respect to the cosine of the angle between the optical axis of the source and normal to the detector
\begin{equation}
I_i = I_m (H,d) \cos \left( \frac{\pi}{2} - \epsilon_s(dt)  \right), 
\end{equation}

where $I_m (H,d)$ is the maximum solar irradiance on a given day at solar noon from \eqref{eqSI}. Thus, the total solar irradiance per unit area $I_{\rm T}$ on a given day and altitude can be estimated by using an approximated elevation angle
time series \cite{arum2020energy}
\begin{equation}
I_{\rm T}(H,d) = \frac{I_{\rm m}(H,d) \tau(\chi,d)}{ \epsilon_{\rm sm}(dt)} \left( 1 - \cos (\epsilon_{\rm sm}(dt))  \right),
\end{equation}

where the daylight time duration $\tau$ at a particular latitude is dependent on the solar azimuthal angle $\Phi(d)$ of the sun and angular distance of the sun $M_a(d)$ as
\begin{equation}
\tau(\chi,d) = 24 \left(1-\frac{1}{\pi}\cos^{-1}\left(\frac{\tan(\chi)\wp(e_o,d)}{\sqrt{1-\wp^2(e_o,d)}}\right)\right),
\end{equation}

with $\wp(e_o,d)={\sin(e_o)\sin(\Phi(d))}$, $M_a(d) = -0.041 + 0.017202d$, and $ \Phi(d) = -1.3411 + M_a(d) + 0.0334 \sin(M_a(d)) + 0.0003 \sin(2M_a(d)) $. The earth is rotating around the tilted axis by an angle of obliquity $e_o = 23.44^o = 0.4093$ with respect to the ecliptic frame \cite{jenkins2013sun}. Moreover, the maximum elevation of the sun $\epsilon_{\rm sm}$ on a given day of the year is a function of latitude in radians $\chi_r$ and the declination angle $\delta$. It can be estimated in radians by using the following equation:
\begin{equation}
\epsilon_{\rm sm} (\chi_r,\delta) = \pi/2 + \chi_r - \delta.
\end{equation}

Thus, the total available solar energy in $kWh$ harvested by the PV cells on a given day can be written as a
\begin{equation}
E_{\rm a} (H,d) = \eta_s  A_p I_{\rm T}(H,d){\rm x} 10^{-3},
\end{equation}

where $\eta_s $ is the power conversion efficiency of solar panels and $A_p = \sum_{s} {A_s}$ is the accumulative solar panel area from $s$ solar panels on HAPS aircraft. Some part of this available energy can be stored in the batteries for NTO and the rest can be expended for station-keeping and communications. Likewise, the instantaneous available solar power at any time of the day can be written as $P_{\rm a} (H,dt) = \eta_s  A_p I(H,dt)$.

\subsection{Aerodynamics}
\begin{figure}[t]
    \centering
   \includegraphics[width=\linewidth]{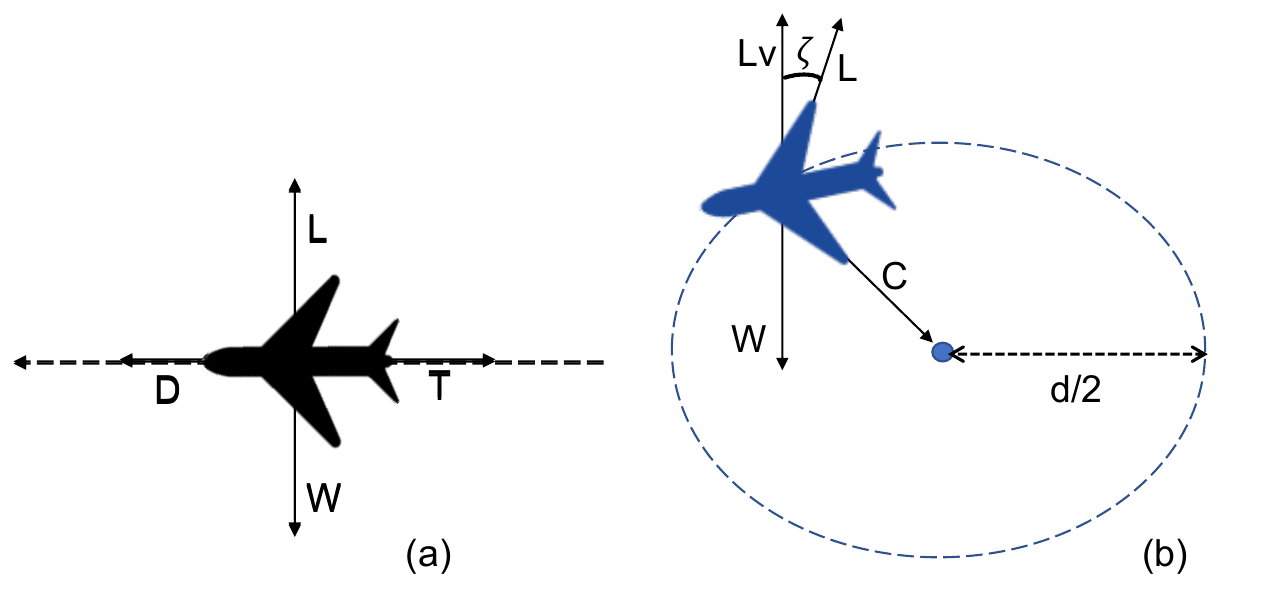}
    \caption{Forces acting upon SHF and SCF }
    \label{fig:SCF}
\end{figure}

In aerodynamics, the steady horizontal flight (SHF) requires a proportional lift force $L$ against the gravitational pull i.e., weight of aircraft $W$ and a thrust $T$ to balance out the drag force $D$ as shown in Fig. \ref{fig:SCF}a. This requires a certain propulsion power $P_{\rm SHF}$ which depends on the true airspeed of HAPS $V$ and the required thrust to maintain a steady-level flight at a given altitude \cite{verstraete2008preliminary}
\begin{equation} \label{eq.P_hs}
  P_{\rm SHF} = T V   / \eta_p \eta_e,
\end{equation}
where $\eta_p$ and $ \eta_e$ are the propeller and engine efficiencies, respectively. The drag is opposed by an equal and opposite thrust as follows:
\begin{equation} \label{eq.Thrust}
T = \frac{1}{2}  \rho_h  V^2  S C_{D_0} + \epsilon \frac{2W^2}{\rho_h S V^2}, 
\end{equation}

where the first part is against the parasitic drag while the second term opposes the induced drag \cite{stengel2015flight}. Moreover, $S$ is the total wing area,  $C_{D_0}$ is the zero-lift drag coefficient and  the coefficient $ \epsilon$  is equal to $\left( \pi e {AR}_w \right)^{-1}$, with Oswald's efficiency factor $e$ and wing aspect ratio $AR_w$. In addition, the air density $\rho_h$ depends on the air pressure $p_h$ and temperature $T_p$ at an altitude $H$ as
\begin{equation}\label{eq.rho}
\rho_h =  \frac{p_h\left(H\right)}{R_{\rm sp}T_p \left( H \right)},
\end{equation}

where ${R_{\rm sp}} $ is the specific gas constant given as $287.052 {\rm J/kg.K}$ and $T_p$ is constant in Stratosphere I but increases by $1$K for every $1$km increase in the altitude for Stratosphere II, as given 
\begin{equation}
T_p \left(H\right) =
\begin{cases} 
T_b,  & 11km \leq H \leq 20km, \\
{T_b+L_b(H-H_{b2} )}.& 20km \leq H \leq 32km. \\
   \end{cases}
\end{equation}

The expression \eqref{eq.P_hs} denotes the power for a SHF. However, we are interested in the steady circular flight (SCF) at the same altitude for station-keeping over a certain area as demonstrated in Fig. \ref{fig:SCF}b. This is primarily achieved by a steady flight with a fixed banking angle $\zeta$ when 
the vertical component of the lift force $L_v$ is equal to $W$ i.e., $L_v = L \cos (\zeta) = W = mg$ and the centripetal force $C$ is equal to $D$ i.e., $L_h = L \sin (\zeta) = C = {2mV^2}/{d}$. Solving these equations simultaneously reveals the required banking angle  
\begin{equation}
\zeta = \tan^{-1}\left( \frac{2V^2}{gd}\right).
\end{equation}

The aircraft load factor is unity for SHF but it is given by $L/W = 1/\cos (\zeta)$ for the SCF. Thus, the propulsion power required $P_{\rm SCF}$ to maintain a SCF at a bank angle $\zeta$ is defined as the product of the square of the load factor and the power required for SHF \cite{arum2020energy}, rendering
\begin{equation}\label{eq.Prop}
P_{\rm SCF} = \left( \frac{1}{\cos (\zeta)} \right)^2 P_{\rm SHF}.
\end{equation}

Furthermore, some additional power is required for the base station payload and avionics named as accessory power $P_{\rm acc}$. Generally, this power is a constant estimation for a given aircraft with certain payload capabilities. 
Hence, the total power requirement for such flight can be computed from the sum of both accessory power and propulsion power i.e., 
\begin{equation}
P_{\rm req} = P_{\rm acc} + P_{\rm pro} =  P_{\rm acc} + P_{\rm SCF}.
\end{equation}
In the presented scenario, the propulsion power required is equal to $P_{\rm SCF}$.
\subsection{Communication Payload}
Ultra-wide terrestrial coverage of the HAPS can be achieved by splitting the service area into multiple cells, as shown in Fig. \ref{fig:HAPS}. Each cell is served by a highly directional beam allowing frequency reuse in the neighboring cells for efficient spectrum allocation. We adopt a polygonal array structure with a downward facing bottom panel and $M-1$ side panels facing outwards. The design includes a beamformer and an array controller which are responsible to create the desired beam and steer it in real time, as detailed in \cite{Nokiahsieh2020uav}. This enables the cell fixation relative to the station-keeping flight pattern for reliable and consistent coverage. 

We assume a multi-cell network, with one center cell and $M-1$ surrounding cells, to be served by a single HAPS. The network consists of
uniformly distributed $K_m$ number of users in the $m^{\rm th}$ cell, where $m \in \{ 1,2,\ldots,M\}$. In each cell coverage, the channel gain is expected to vary with the distance from the center as well as on the azimuth angle. The strongest channel gain is available at the center of the cell along the boresight direction $\theta = 0$. However, as the distance varies and/or the azimuth direction deviates from the boresight, the performance can be degraded due to the increased path loss, lower radiation pattern gain, and higher co-channel interference (CCI) from the neighboring cells. The striking difference in the channel gains of the users in each cell enables us to reap maximum benefits offered by NOMA. Consider the DL-NOMA scenario, where $m^{\rm th}$ cell is served by a directional beam with superposition coding as \footnote{The transmitted/received signals, channel gains and allocated powers are function of time. However, the time notation is omitted for brevity.}
\begin{equation}
x_m = \sum_{k=1}^{K_m}{\sqrt{P_m \alpha_k^m} s_k^m },
\end{equation}

where $P_m $ is the power budget for $m^{\rm th}$ cell, whereas $\alpha_k^m$ and  $s_k^m$ are the fraction of power allocated to and intended information signal for the $k^{\rm th}$ user in $m^{\rm th}$ cell, respectively. It is important to highlight that $\sum_{k=1}^{K_m}{ \alpha_k^m} = 1, \; \forall m,$ in order to limit the power division within given budget. 
The presented system undergoes two types of interference: 
\begin{itemize}
\item Intra-cell interference (IACI) owing to NOMA 
\item Inter-cell interference (IECI) due to frequency reuse
\end{itemize}

Therefore,  using the conventional wireless communication model, the received signal at user $l$ in the $m^{\rm th}$ cell is given by 
\begin{align}
y_l^m = & \underbrace{h_{lm}^m{\sqrt{P_m \alpha_l^m} s_l^m }}_{\text{Desired Signal}}   +  \underbrace{h_{lm}^m \sum_{k=1 \atop k \neq l}^{K_m}{\sqrt{P_m \alpha_k^m} s_k^m}}_{\text{IACI}} +   \nonumber \\ & \qquad \qquad
+\underbrace{ \sum_{j=1 \atop j \neq m}^{J_m} h_{lm}^j \sum_{k=1}^{K_j}{\sqrt{P_j \alpha_k^j} s_k^j }}_{\text{IECI}} + w_l^m,
\end{align}

where, $h_{lm}^j$ is the channel gain coefficient between $j^{\rm th}$ HAPS panel and $l^{\rm th}$ user in $m^{\rm th}$ cell and $w_l^m$ is the receiver thermal noise modeled as circular symmetric complex Gaussian random variable,
i.e., $w_l^m \sim  \mathcal{CN}(0,\sigma^2)$. Moreover, $J_m$ is the set of neighbouring interfering cells around $m^{\rm th}$ cell e.g., $J_1$ is the set of all $M-1$ cells surrounding the center cell, whereas $J_{ \lbrace m\backslash 1 \rbrace}$ is the set of $3$ immediate neighbors around $m^{\rm th}$ edge cell. 

\subsection{Energy Storage}
The consistent flight operation of HAPS for long-endurance requires the energy storage of the surplus power for NTO. The electrical energy is stored and drawn from the rechargeable batteries depending on the net power $P_{\rm net}$ i.e.,
\begin{equation} \label{eq.netP}
P_{\rm net} = P_{\rm a} - P_{\rm req} - P_{\rm T},
\end{equation}

where $P_{\rm net}$ is the power remainder from the available solar power $P_{\rm a}$ after meeting all the aerodynamic $P_{\rm req} $ and communication requirements $P_{\rm T} = \sum_m P_m $. The battery energy moves from state $E_b^{i-1}$ to state $E_b^{i}$ either by charging or by discharging as  \cite{FBmarriott2020trajectory}
\begin{equation}
E_b^i = E_b^{i-1} + \eta_b P_{\rm net}^i \Delta t, 
\end{equation}

where  battery efficiency $\eta_b$ is either charging $\eta_c$  or discharging $\eta_d$ efficiency and is given by 
\begin{equation}
\eta_b =
\begin{cases} 
\eta_c,  &  P_{\rm net} \geq  \mu, \\
\eta_d. & P_{\rm net} \leq \mu, \\
   \end{cases}
\end{equation}

 with $\mu$ as the minimum power required to charge the batteries and $\Delta t $ is the time separation between two battery states when the net power is almost constant.

To summarize, energy budgeting dictates the ability of a certain HAPS to harvest, store, and retrieve the abundant solar energy as electrical energy. This electrical energy is then consumed by the electric motors and propellers for propulsion, avionics, payload and communication equipment. The aim of this paper is to limit the power expenses within power budget for a self-sustaining operation. The next section is dedicated for the communication payload and describes the suitable propagation model for accurate performance analysis. 

\section{Propagation Model and Link Budget}
\label{SecIII}
The radio signal propagation from HAPS to the UE undergoes free space path loss and multipath fading due to the significant distance between them and obstacles around the UE, respectively\footnote{Note that the HAPS station-keeping flight does not contribute to the fast fading since there is no moving scatter surrounding the aircraft \cite{Nokiahsieh2019propagation}. }. Therefore, the propagation loss of the adopted system is modeled as a combination of both small-scale and large-scale fading. Hence, the channel coefficient $h_{lm}^j$ can be expressed as follows:
\begin{equation}\label{eq.Link}
h_{lm}^j = \frac{g_{lm}^j \sqrt{ G^j_m}}{\sqrt{L(d_l^m)}}, 
\end{equation}

where $g_{lm}^j $ is the small scale fading coefficient between the $j^{\rm th}$ transmitting panel and $l^{\rm th}$ user in $m^{\rm th}$ cell, 
$G^j_m$ is the array gain for the link between $j^{\rm th}$ panel and  $m^{\rm th}$ cell, and $L(d_l^m)$ is the path loss as a function of $d_l^m$ i.e., the distance between HAPS and $l^{\rm th}$ user in $m^{\rm th}$ cell. The computation of these parameters is highlighted in the subsequent sections.

\subsection{Small Scale Multi-path Fading}
The received signal comprises of both the Line-of-Sight (LOS) and Non Line-of-Sight (NLOS) components pertaining to the HAPS bore-sight position and independent diffuse multipath reflections from the obstacles. The LOS component is generally deterministic whereas the envelope of NLOS component is modeled as a Rayleigh random variable. Hence, the aggregate small-scale multipath fading coefficient $g_{lm}^j$ is modeled as a Rician distributed random variable \cite{cuevas2004channel,kanatas2017radio,oladipo2007stratospheric}
\begin{equation}
f(x\mid \nu ,\sigma_{\rm f })={\frac  {x}{\sigma_{\rm f } ^{2}}}\exp \left({\frac  {-(x^{2}+\nu ^{2})}{2\sigma_{\rm f } ^{2}}}\right)I_{0}\left({\frac  {x\nu }{\sigma_{\rm f } ^{2}}}\right),
\end{equation}

where $I_0$ denotes zeroth-order modified Bessel function of the first kind whose shape parameter is defined by the ratio between the average power of LOS component and the average power associated with NLOS multipath components i.e., ${{ {\nu ^{2}}/{2\sigma_{\rm f } ^{2}}}}$.
\subsection{Directivity Gain}
The communication panels are equipped with antenna arrays which are responsible for directional beamforming. The sectorial antenna pattern is favorable for the minimal CCI and yields the following array gain \cite{sun2018performance}
\begin{equation}
G^j_m \left(\theta_{lm}^j \right) =
\begin{cases} 
M_b,   & |\theta_{lm}^j| \leq \theta^b \\
m_b,   & \text{otherwise} \\
   \end{cases}
\end{equation}

where $\theta_{lm}^j$ is the angle of departure and $\theta^b$ is the half power beamwidth of the main lobe. Moreover, the directivity gains of main lobe and back lobe are denoted by $M_b$ and $m_b$, respectively. Evidently, the relation $m_b << M_b$ holds due to the decreasing antenna gain while moving away from the boresight position in a horizontal direction. Hence, the sectorial antenna pattern reaps the maximal directional gain rendering the minimal IECI to the users in neighboring cells.  
\subsection{Link Budget}
The large scale propagation is characterized as a free space path loss model with the direct distance $d_l^m$ between HAPS and ${\rm UE}_l^m$ as $d_l^m = H/\sin \psi_l^m$, where $\psi_l^m$ is the elevation angle of HAPS from  ${\rm UE}_l^m$. 
Evidently, the users in the center cell enjoy a larger elevation angle between $\psi_c \leq \psi_l^1 \leq \pi/2$, whereas the edge cell users are at relatively smaller elevation angles $\psi_e \leq \psi_l^{\lbrace m\backslash 1 \rbrace} \leq \psi_c$, with $0 \leq \psi_e \leq \psi_c \leq \pi/2$, where $\psi_c$ and $\psi_e $ depend upon the radius of center cell and entire coverage area, respectively. We employ the space communication model for the aerial HAPS to compute the received signal path loss $L(d_l^m)$ as \cite{Space2017}
\begin{equation}
L(H, \psi_l^m) = \frac{16\pi^2 H^2 }{\lambda^2 \sin^2 (\psi_l^m)},
\end{equation}

where $\lambda$ is the wavelength corresponding to the carrier frequency. The free-space path loss is computed as the ratio between transmit power and received power. The  assumption of isotropic receiver antenna, in the DL-NOMA communication, renders an effective receiver area $\lambda^2 / 4\pi$ and the received signal intensity is given by the inverse-square law. 
\subsection{Rate Analysis}
Consider the $K_m$ users are uniformly distributed and ordered as $u_1^m, u_2^m, \ldots, u_{K_m}^m$ depending on their decreasing IECI and increasing channel strengths. Given this ordered arrangement and successive interference cancellation (SIC) at user $u_l^m$, it is capable of decoding all users from $u_1^m$ to $u_{l-1}^m$ and subtracting these from the received signal. Thus, it can decode its own signal from the resultant by considering the interference from $u_{l+1}^m$ to $u_{K_m}^m$ as noise. Moreover, the IECI in a given cell is also treated as noise at all users. Therefore, the signal-to-interference noise ratio $\gamma$ at user $u_l^m $ is given by
\begin{equation}\label{eq.SINR1}
\gamma_l^m = \frac{|h_{lm}^m|^2{{P_m \alpha_l^m}}}{|h_{lm}^m|^2 \sum_{k= l+1}^{K_m}{{P_m \alpha_k^m}  }+ {\mathcal{I}}_l^m+\sigma^2},
\end{equation}

where the noise power is given as ${\sigma^2}({\rm dBm}) = -174 + 10\log(B)+{\rm NF}$ with NF denoting the noise figure \cite{shibata2020system} and the inter-cell interference power is given by
\begin{equation}
 {\mathcal{I}}_l^m =  \sum_{j=1 \atop j \neq m}^{J_m} \sum_{k=1}^{K_j} |h_{lm}^j|^2 {{P_j \alpha_k^j}} = \sum_{j=1 \atop j \neq m}^{J_m} |h_{lm}^j|^2 {{P_j}}.
\end{equation}

Assuming perfect decoding with perfect receiver CSI and user ordering, the achievable rate of user $u_l^m $ is
\begin{equation}
R_l^m =B \log_2 \left[ 1+{\gamma}_l^m  \right].
\end{equation}

The sum rate $R$ of all users in all $M$ cells can be written as $R = \sum_{m= 1}^{M} R_m$, where the sum rate of users in $m^{\rm th }$ cell is given by $R_m = \sum_{l= 1}^{K_m}{ R_l^m}$ yielding
\begin{equation}
R =\sum_{m= 1}^{M} R_m =B \sum_{m= 1}^{M}\sum_{l= 1}^{K_m} \log_2 \left[ 1+ {  \gamma}_l^m   \right].
\end{equation}

$ {  \gamma}_l^m $ in \eqref{eq.SINR1} can be expressed using \eqref{eq.Link} as
\begin{equation}\label{eq.SINR2}
 {\gamma}_l^m \!\! = \! \frac{|g_{lm}^m|^2M_b \alpha_l^m}{|g_{lm}^m|^2 M_b \!\!\sum\limits_{k= l+1}^{K_m}{\alpha_k^m}\!\!+\!\! m_b \sum \limits_{j= 1}^{J_m}\frac{P_j}{P_m}|g_{lm}^j|^2 \!\! +\!\!\frac{ L(H,\psi_l^m)}{\varrho_m} }, 
\end{equation}

where $\varrho_m$ is the transmit SNR for $m^{\rm th}-$cell i.e., $\frac{P_m}{\sigma^2}$. User experiences high IACI and low IECI owing to high directivity gain $M_b$ and low antenna gain $m_b$ for the users outside the main lobe, respectively. 
We can now formulate the optimization problem to design optimal power allocation in order to maximize the sum rate of all users within the allocated power budget.
\section{Problem Formulation}
This work emphasizes the joint optimization of flight and communication parameters given inevitable power limitations.  We aim to design the optimal altitude and UAV airspeed (in a circular trajectory of fixed radius) for the station-keeping flight along with the optimal power allocation for all the DL-NOMA users in its coverage area.  
We formulate two different design problems for day and night operation based on their different characteristics:
\begin{enumerate}
\item The availability of abundant solar power during the day allows optimal flight and maximum transmission power while storing an essential amount of solar energy for NTO.
\item The limited stored energy should support the dwell flight for entire night while satisfying a minimum data-rate constraint. 
\end{enumerate}
\subsection{Day Time Operation}
The objective of DTO is to maximize the sum rate of all users after storing the adequate energy for night time. We divide the daylight period into $n$ equal time intervals (each with almost constant solar flux) and then optimize the power allocation $\alpha[n]= \{\alpha_1,\alpha_2, \ldots,\alpha_m \}$ with $\alpha_m = \{\alpha_1^m,\alpha_2^m, \ldots,\alpha_{K_m}^m \}$, altitude $H[n]$ and airspeed $V[n]$ for each interval. 
\begin{subequations}\label{eq.P1}
\begin{alignat}{2}
\textbf{P1}:\quad &\!\!\!\!\underset{\alpha[n], H[n],V[n] }{\text{maximize }}
&&{\sum_{m= 1}^{M}   \sum_{l = 1}^{K_m}  R_l^m }\left( {\alpha[n]} ,H[n] \right) \\
&\!\!\!\! \text{subject to} \quad
& &   R_l^m \geq \Omega_m, \, \forall l,m,   \label{eq.p3}   \\ 
& & &  {  \sum\nolimits_{k = 1}^{K_m}  \alpha_k^m[n] } = 1,  \quad \forall m \label{eq.p1} \\
& &&  0 \leq{  \alpha_k^m[n] } \leq 1, \quad \forall k,m  \label{eq.p2} \\
&&& \alpha_1^m[n] \geq \alpha_2^m[n] \ldots \geq \alpha_{K_m}^m[n], \, \forall m,  \label{eq.p2.} \\
& & &  P_{\rm net} (H[n],V[n]) \geq   P_{\rm st}, \label{eq.p4} \\
& & & \sum\nolimits_{m = 1}^{M} {P_m}[n] \leq \Upsilon P_{\rm T}, \label{eq.p5} \\
& & & H_{\rm min} \leq H[n] \leq H_{\rm max},  \label{eq.p6} \\
& & &  V_{\rm min}  \leq V[n] \leq  V_{\rm max},  \label{eq.p7} 
\end{alignat}
\end{subequations}
where the QoS rate constraint \eqref{eq.p3} and the sum power constraint \eqref{eq.p1} ensure the minimum achievable rate $\Omega_m$ for each user in $m^{\rm th}$ cell and the transmission within power budget, respectively.
Moreover, the boundary constraints on power, altitude, and airspeed are given by \eqref{eq.p2},
\eqref{eq.p6} and \eqref{eq.p7}, respectively. It is important to highlight that $H_{\rm min}$, $ H_{\rm max}$ and $ V_{\rm max}$ are constants but the stalling speed $ V_{\rm s}$ is a function of aircraft dimensions and altitude as $ V_{\rm s} = \sqrt{{2W}/{\rho_h(H)S C_{\rm Lmin}}}$. It is the minimum speed $V_{\rm min} $ at $H$ to maintain a steady-level flight. In addition, the constraint \eqref{eq.p4} ensures the availability of requisite power stored $P_{\rm st}$ in batteries for NTO. Finally, \eqref{eq.p5} restricts the communication power budget for all cells after accounting the feed line losses $\Upsilon$ encountered while drawing effective radiated power from the actual transmission power. 

The problem $\textbf{P1}$ is jointly non-convex in the given optimization variables, therefore we propose alternate optimization (AO) and solve the sub-problems \eqref{eq.P1a}, \eqref{eq.P1b} and \eqref{eq.P1c} iteratively as presented in Algorithm \ref{Algo1}. We initialize resource allocation by selecting a time and date at instance $n$ and choosing the QoS threshold $\Omega_m$. Feasible starting altitude and speed are used to compute the required propulsion power $P_{\rm pro}[n]$ which is then subtracted from the available power at that instance $P_{\rm a}[n]$ along with the $P_{\rm st}$ to obtain the transmission power $P_{\rm T}[n]$. Then, subproblems \textbf{P1(a)}, \textbf{P1(b)}, and \textbf{P1(c)} are solved iteratively using the results of previously solved problems. Next, we update $P_{\rm T}[n]$ using the renewed $P_{\rm pro}[n]$ to find the sum-rate $R[n]$. Alternate optimization works in an iterative manner till it meets the stopping criteria to furnish the optimum values of the $\alpha[n]$, $H[n]$ and $V[n]$.
\begin{algorithm}[!t]
\caption{Alternate Optimization: Day Time Operation }\label{Algo1}
\begin{algorithmic}[1]
\State \textbf{Initialize}  ${R}^{(i-1)}\gets 0$, $i \gets 1$,  and  $\epsilon \gets \infty$   \textbf {Set} tolerance $\delta$
\State \textbf{Select} a time instance $n$ and QoS Thresholds $\Omega_m$.
\State \textbf{Choose} feasible starting points $H[n]^{(i-1)}$ and $V[n]^{(i-1)}$.
\State Compute $P_a[n]$ using $j_d(n)$, $\epsilon_s(n)$, and $H[n]^{(i-1)}$
\State Determine $P_{\rm pro}[n]^{(i-1)}$ using $H[n]^{(i-1)}$ and $V[n]^{(i-1)}$
\State Evaluate $P_{\rm T}[n]^{(i-1)}$ by subtracting $P_{\rm pro}[n]^{(i-1)}$ in \eqref{eq.netP}
\While {$\epsilon \ge \delta$}
\State  Solve \textbf{P1(a)} using  $H[n]^{(i-1)}$, $P_{\rm T}[n]^{(i-1)}$ to  get ${\alpha[n]}^{(i)}$ 
\State Solve \textbf{P1(b)} using  ${\alpha[n]}^{(i)}$ to obtain $H[n]^{(i)}$
\State Solve \textbf{P1(c)} using  $H[n]^{(i)}$ to obtain $V[n]^{(i)}$
\State Calculate $P_{\rm pro }[n]^{(i)}$ using $  H[n]^{(i)}$ and $V[n]^{(i)}$
\State Find $P_{\rm T}[n]^{(i)}$  from $P_a[n]$ deducting $P_{\rm pro }[n]^{(i)}$ and $P_{\rm st}$ 
\State Evaluate  $R^{(i)}$ using ${\alpha[n]}^{(i)}$, $H[n]^{(i)}$ and $P_{\rm T}[n]^{(i)}$
\State Update  ${\epsilon \gets \left\| {R}^{(i)}- {R}^{(i-1)}\right\|}$ and  ${i \gets i+1}$ 
\EndWhile
\State   $R^*[n] \gets R[n]^{(i)}$, $\alpha^*[n] \gets \alpha[n]^{(i)}$, $H^*[n] \gets H[n]^{(i)}$ and $V^*[n] \gets V[n]^{(i)}$

\end{algorithmic}
\end{algorithm}

For a given flight at altitude $H$ with airspeed $V$ at an instant $n$, the power allocation problem is given by \eqref{eq.P1a}. The transmit power budget is calculated from  \eqref{eq.netP} after deducting $P_{\rm req}[n]$ from the available solar power $P_a[n]$ for a fixed $P_{\rm st}$.
\begin{subequations}\label{eq.P1a}
\begin{alignat}{2}
\textbf{P1(a)}:\quad &\!\!\!\!\underset{\alpha[n] }{\text{maximize }}
&&{\sum_{m= 1}^{M}   \sum_{l = 1}^{K_m}  R_l^m }\left( {\alpha[n]} ,H[n] \right) \\
&\!\!\!\! \text{subject to} \quad
& &   \eqref{eq.p1},\eqref{eq.p2},\eqref{eq.p3},\eqref{eq.p5}.  
\end{alignat}
\end{subequations}
Interestingly, the optimization problem \textbf{P1(a)} can be solved independently in $\alpha_1, \alpha_2,\ldots,\alpha_m$ as it is a disjoint problem for all cells, given a fixed altitude and equal power budget i.e., $P_m[n] = \Upsilon P_{T}/M$. The elaborated closed-form solution to this problem is presented in \cite{wang2019user} as:

\begin{theorem}
In the considered power allocation problem, the sum rate and minimum power coefficients of users in $m^{\rm th}-$cell, respectively, are given by
\begin{equation}
R_m = K_m \Omega_m + B \log_2 \left[ 1+  \frac{1 - \sum\nolimits_{k= 1}^{K_m} \hat{\alpha}_l^m }{ A_{K_m}^m  } \right],   
\end{equation}
\begin{equation}\label{eq.PowerAlloc}
\hat{\alpha}_l^m = \left(2^{\Omega_m'}-1 \right) \left( \sum\limits_{k= l+1}^{K_m}{\hat{\alpha}_k^m} +  A_{l}^m  \right),
\end{equation}
where
\begin{equation}
A_{l}^m =  \frac{m_b}{M_b {P_m|g_{lm}^m|^2}} \sum \limits_{j= 1}^{J_m}{P_j |g_{lm}^j|^2} \!\! +\!\!\frac{ L(H,\psi_l^m)}{\varrho_m M_b|g_{lm}^m|^2}. 
\end{equation}
if the following condition holds
\begin{equation}
\left(  2^ {\Omega_m' }-1 \right)   \left(  \sum\limits_{i= 1}^{K_m}  2^{(i-1)\Omega_m' } A_{i}^m  \right) \leq 1.
\end{equation}
\end{theorem}
 The first term of $R_m$ is the QoS thresholds of all users in $m^{\rm th}-$cell whereas the second term is the additional rate of $K_m$ user \footnote{It is important to highlight that the users are ordered with decreasing $A_{l}^m $ i.e., $A_{1}^m \geq A_{2}^m \geq \ldots \geq A_{K_m}^m  $.   } after allocating the remaining power  $1-\sum\nolimits_{k= 1}^{K_m} \hat{\alpha}_l^m$ to it in order to maximize the sum rate. The normalized rate target $\Omega_m'$ is obtained as $\Omega_m / B$.

\begin{theorem}
For $\sum\limits_{k= 1}^{K_m} \hat{\alpha}_l^m \geq 1 $, there exists a user $u$ in $1\leq u \leq K_m$ which satisfies the following condition:
\begin{equation}
\begin{cases} 
\left(  2^ {\Omega_m' }-1 \right)   \left(  \sum\limits_{i= u+1}^{K_m}  2^{i-1} A_{i}^m     \right)  \leq 1, &  \\
\left(  2^{ \Omega_m'} -1 \right)   \left(  \sum\limits_{i= u}^{K_m}  2^{i-1} A_{i}^m     \right) \geq 1. & \\
   \end{cases}
\end{equation}
Hence, the maximum achievable sum rate is given by
\begin{equation}
R_m = \left(K_m-u\right) \Omega_m' + \log_2 \left[ 1+  \frac{\Delta \alpha}{    1-\Delta \alpha +     A_{u}^m  } \right],   
\end{equation}
where 
\begin{equation}
\Delta \alpha = 1-\left(  2^ {\Omega_m'} -1 \right)   \left(  \sum\limits_{i= u+1}^{K_m}  2^{i-1} A_{i}^m     \right).
\end{equation}
\end{theorem}
The first term of $R_m$ is the QoS thresholds of users from $u+1$ to $K_m$ and the second term is the rate of user $u$ with power $\Delta \alpha$. It signifies that only $u+1$ to $K_m$ can attain QoS threshold with powers $\hat{\alpha}_{u+1}^m, \hat{\alpha}_{u+2}^m, \ldots, \hat{\alpha}_{K_m}^m$, respectively, using \eqref{eq.PowerAlloc}. However, the users before (and including) $u^{\rm th}$ user cannot achieve their target rates. So, the remaining power $\Delta \alpha$ is allocated to the $u^{\rm th}$ user.
The next optimization problem maximizes the sum rate of all users with respect to altitude for a given power allocation $\alpha[n]$.
\begin{subequations}\label{eq.P1b}
\begin{alignat}{2}
\textbf{P1(b)}:\quad &\!\!\!\!\underset{H[n] }{\text{maximize }}
&&{\sum_{m= 1}^{M}   \sum_{l = 1}^{K_m}  R_l^m }\left( {\alpha[n]} ,H[n] \right) \\
&\!\!\!\! \text{subject to} \quad
& &   \eqref{eq.p4},\eqref{eq.p6}.  
\end{alignat}
\end{subequations}
This problem is equivalent to minimizing $L(H, \psi_l^m) $ subject to \eqref{eq.p4} and \eqref{eq.p6}. Evidently, $L(H, \psi_l^m) $ is strictly increasing in $H$ and the solution is one word i.e., 
$H^* [n]= H_{\rm min}$ while satisfying $ \eqref{eq.p4}$.

\begin{subequations}\label{eq.P1c}
\begin{alignat}{2}
\textbf{P1(c)}:\quad &\!\!\!\!\underset{V[n] }{\text{maximize }}
&&{\sum_{m= 1}^{M}   \sum_{l = 1}^{K_m}  R_l^m }\left( {\alpha[n]} ,H[n] \right) \\
&\!\!\!\! \text{subject to} \quad
& &    \eqref{eq.p4},\eqref{eq.p7}.  
\end{alignat}
\end{subequations}
This problem optimizes airspeed $V[n]$ at a given altitude $H[n]$. 
Although the sum rate is not directly affected by the airspeed, it is a function of the available transmission power which is dictated by the propulsion power. Therefore, in order to provide the large transmission power, it is imperative to choose $V[n]$ which decreases the propulsion power. The solution to such a problem is detailed in Section \ref{ssec:NTO}.

%
%
%
\begin{algorithm}[!t]
\caption{Alternate Optimization: Night Time Operation }\label{Algo2}
\begin{algorithmic}[1]
\State \textbf{Initialize}  $P_{\rm pro}^{(i-1)}\gets 0$, $i \gets 1$,  and  $\epsilon \gets \infty$   
\State \textbf {Set} tolerance $\delta$ and \textbf{Select} a time instance $n$ 
\State \textbf{Choose} feasible starting point $H[n]^{(i-1)}$ 
\While {$\epsilon \ge \delta$}
\State  Evaluate $V[n]^{(i)}$  using $H[n]^{(i-1)}$ in \eqref{eq.Vstar} 
\State  Compute $H[n]^{(i)}$  using $V[n]^{(i)}$ in \eqref{eq.Hstar} 
\State Calculate $P_{\rm pro}^{(i)}$  using $H[n]^{(i)}$ and  $V[n]^{(i)}$
\State Update  ${\epsilon \gets \left\| P_{\rm pro}^{(i)}- P_{\rm pro}^{(i-1)}\right\|}$ and  ${i \gets i+1}$ 
\EndWhile
\State   $P_{\rm pro}^*[n] \gets P_{\rm pro}[n]^{(i)}$, $H^*[n] \gets H[n]^{(i)}$, $V^*[n] \gets V[n]^{(i)}$
\end{algorithmic}
\end{algorithm}

\begin{figure}[t]
    \centering
   \includegraphics[width=0.85\linewidth]{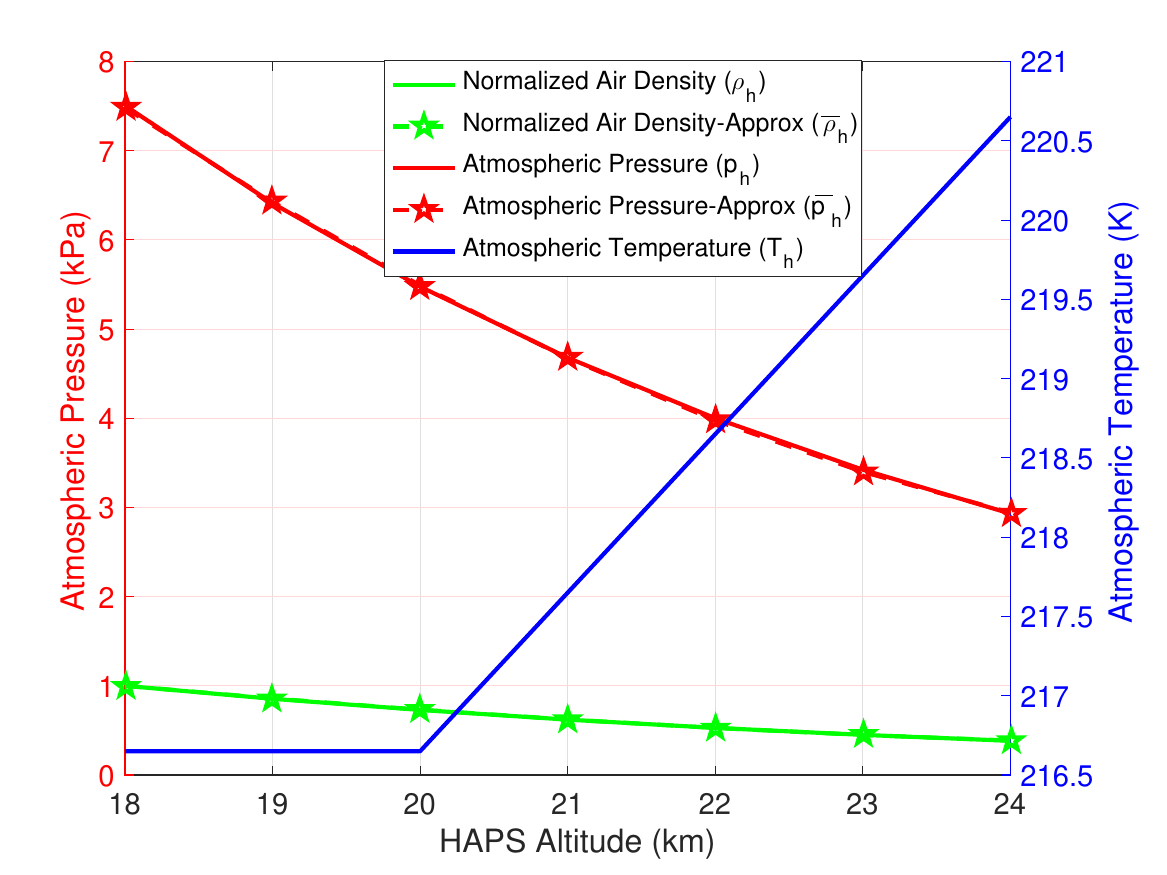}
    \caption{Atmospheric Pressure and Temperature with varying altitude in Stratosphere. }
    \label{fig:APT}
\end{figure}
\subsection{Night Time Operation}\label{ssec:NTO}
The night time function of HAPS is largely limited by the stored power in rechargeable batteries. Therefore, we propose a flight which minimizes the power consumption during night time and transmission with a fixed power budget. We formulate \textbf{P2} to minimize the propulsion power requirement with respect to $H$ and $V$.
\begin{subequations}\label{eq.P2}
\begin{alignat}{2}
\textbf{P2}:\quad &\!\!\!\!\underset{ H[n],V[n] }{\text{minimize }}
&&    { P_{\rm pro} } ( V[n], H[n])    \\
&\!\!\!\! \text{subject to} \quad
& &     \eqref{eq.p6},   \eqref{eq.p7}.    \label{eq.p24} 
\end{alignat}
\end{subequations}
The problem \textbf{P2} is jointly non-convex in optimization variables $H$ and $V$ pertaining to the indefinite Hessian matrix. Nonetheless, it is convex in both $H$ and $V$ \footnote{The proof of indefinite $\nabla^2 P_{\rm pro}$ individually for a fixed banking angle and non-negative second-order gradients
${\partial^2 P_{\rm pro}}/{\partial H^2} $ and ${\partial^2 P_{\rm pro}}/{\partial V^2}$ is straight-forward but omitted for brevity.}.  
Hence, we breakdown \textbf{P2} to \textbf{P2(a)} and \textbf{P2(b)} in order to minimize the propulsion power requirement  with respect to $ V[n]$ and $ H[n]$, respectively. 
\begin{subequations}\label{eq.P2a}
\begin{alignat}{2}
& \textbf{P2(a)}:\quad &&\!\!\!\!\underset{V[n] }{\text{minimize }}
   { P_{\rm pro} } ( V[n], H[n])     \quad
 \text{s.t.} \quad
  \eqref{eq.p7}.  \\
&\textbf{P2(b)}:\quad &&\!\!\!\!\underset{H[n] }{\text{minimize }}
    { P_{\rm pro} } ( V[n], H[n])    
 \quad \text{s.t.} \quad
  \eqref{eq.p6}.    
\end{alignat}
\end{subequations}
The propulsion power in \eqref{eq.Prop} is convex and strictly increasing in $V$. 
Therefore, we can find the closed-form optimal solution of \textbf{P2(a)} as
\begin{equation}\label{eq.Vstar}
V^*[n]=
\begin{cases} 
V_m[n],  &    V_s[n] \leq V_m[n] \leq V_{\rm max}, \\
V_s[n] ,  & V_m[n] \leq  V_s[n].\\
   \end{cases}
\end{equation}
where $V_m[n]$ is derived using the second order sufficient optimality condition.  
\begin{equation}\label{eq.Vm}
   V_m [n]   = \sqrt {  \frac{2W}{\rho_h[n]  S } \sqrt{\frac{\epsilon}{3C_{D_0}}      }}. 
\end{equation}
Evidently, $V_m[n]$ is the global optimal solution of \textbf{P2(a)} if it is in the desired range  \eqref{eq.p7}, otherwise $V_s[n]$ is the local minimum below which steady level flight is not possible. 
Similarly, $P_{\rm pro}$ is a convex and strictly decreasing function of altitude $H$, enabling us to compute the optimal solution of \textbf{P2(b)} as
\begin{equation}\label{eq.Hstar}
 H^*[n]=
\begin{cases} 
H_m[n],  &  H_{\rm min} \leq H_m[n] \leq  H_{\rm max}\\
H_{\rm max} ,  & \text{otherwise}\\
   \end{cases}
\end{equation}

 where $H_m[n]$ is the stationary point and obtained using the second order sufficient condition
\begin{equation}\label{eq.Hm}
H_m[n]=
\begin{cases} 
H_{b1}   - \frac{{RT_{b}}}{g M} \ln  \left( \frac{\varpi[n] T_b }{p_{b1} }  \right),  & \text{Stratosphere-I} \\
H_{b2} - \frac{T_b}{L_b} + \frac{1}{L_b} \left(  \frac{p_{b2}   T_b^{\frac{g M}{RL_b}}}{\varpi[n] }  \right)^\vartheta ,  & \text{Stratosphere-II}  \\
   \end{cases}
\end{equation}

where $\varpi[n] = \frac{2WR_{\rm sp}}{S  V[n]^2}\sqrt{\frac{\epsilon}{C_{D_0}}}$ and $\vartheta = {\frac{{RL_b}}{{g M+RL_b}}}$. We either choose the feasible case or Stratosphere-II if both are feasible. Again, $H_m[n]$ is the global solution to \textbf{P2(b)} if it satisfies \eqref{eq.p6} otherwise $ H_{max}$ is the local optimal owing to the strictly decreasing nature of $P_{\rm pro}$.

\begin{figure*}[!t]
 \begin{minipage}[b]{0.333\textwidth}
  \centering
 {{\includegraphics[width=\linewidth]{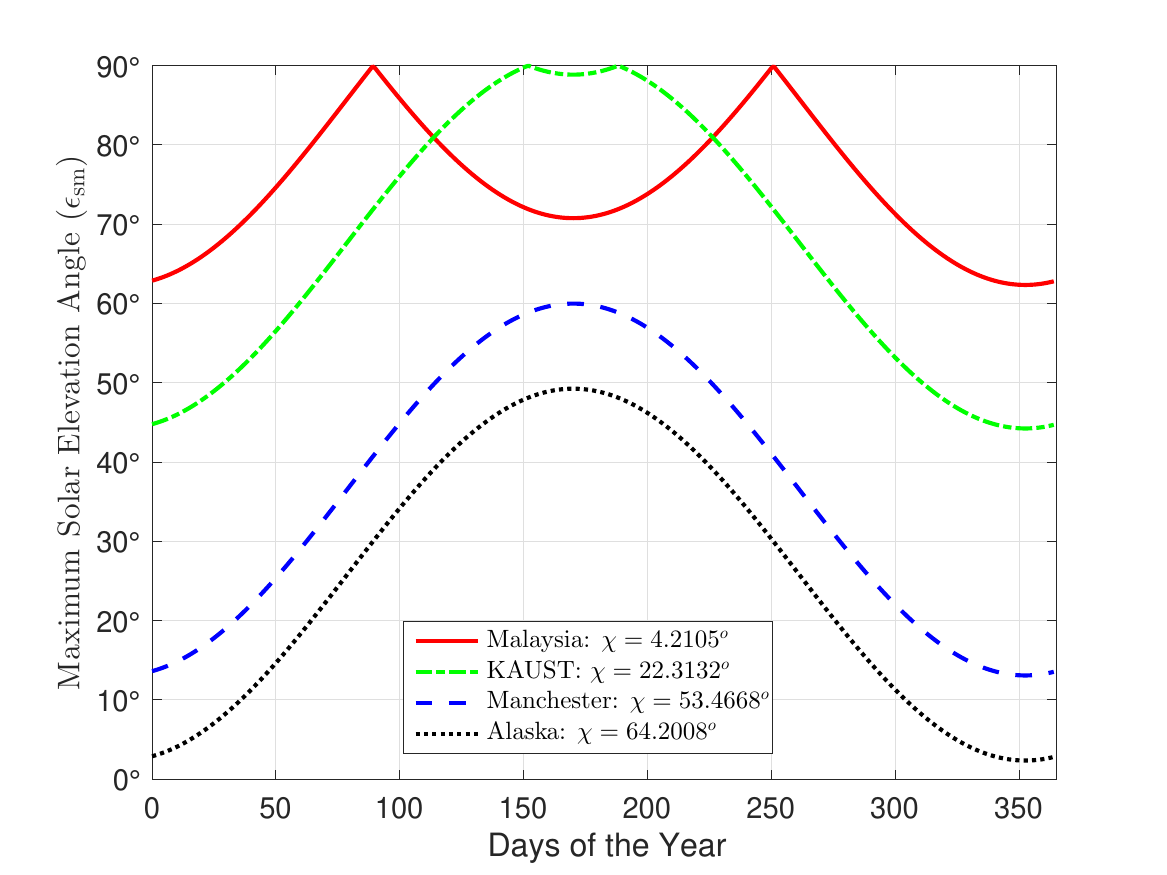}}}
\end{minipage}
\begin{minipage}[b]{0.333\textwidth}
  \centering
 {{\includegraphics[width=\linewidth]{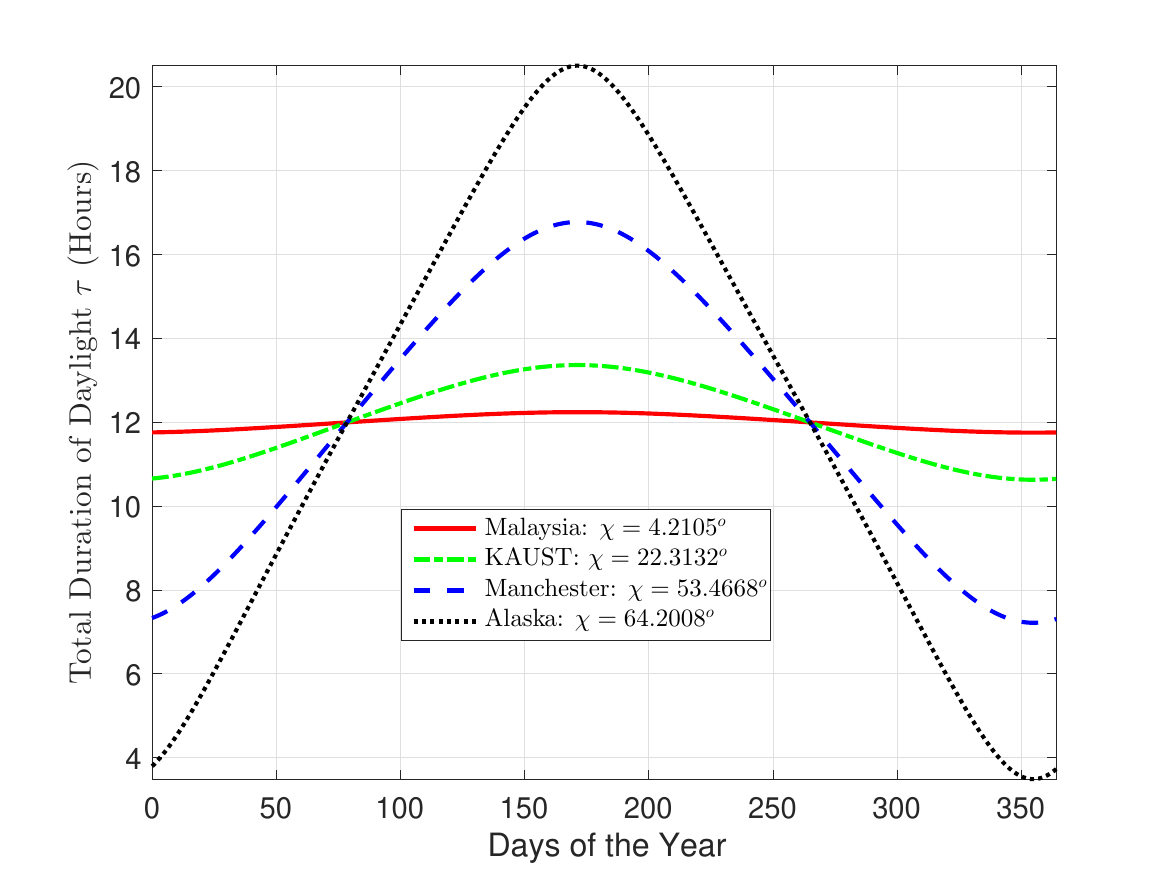}}}
\end{minipage}
\begin{minipage}[b]{0.333\textwidth}
  \centering
{\includegraphics[width=\linewidth]{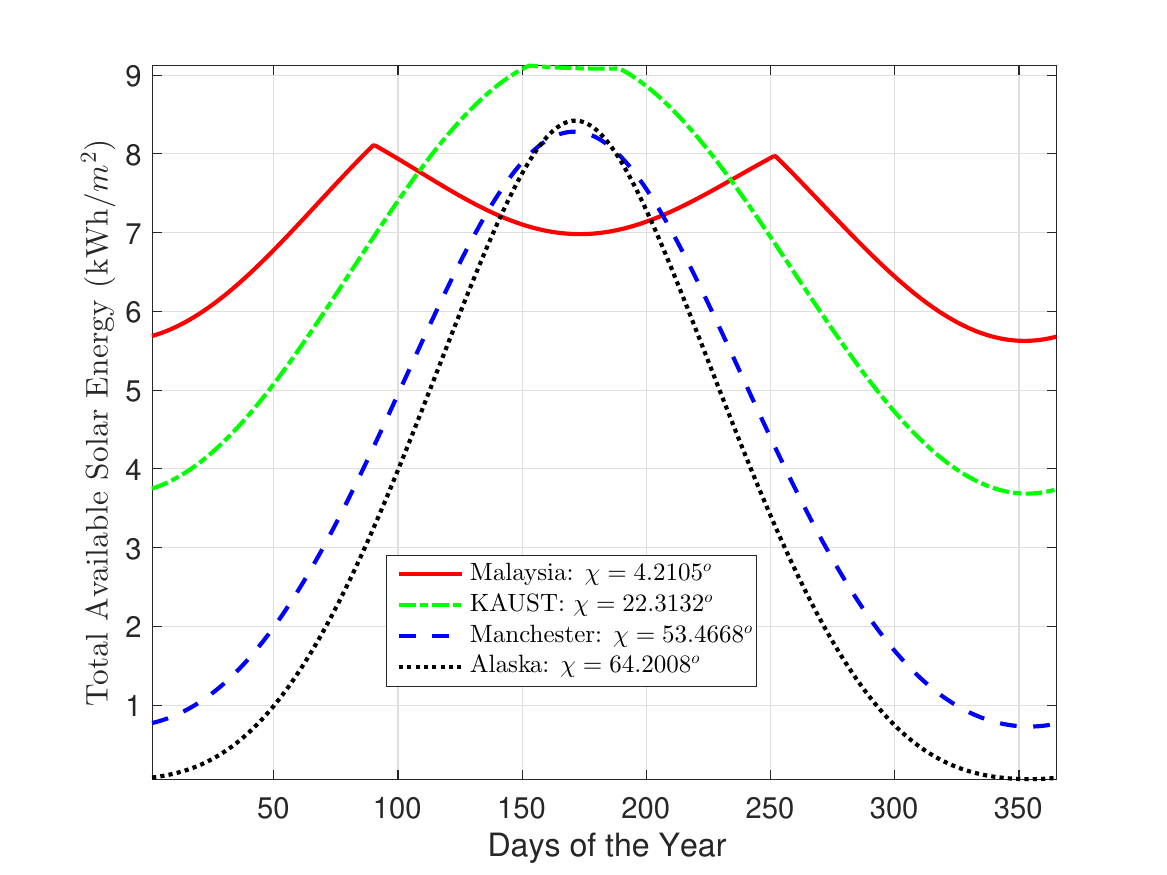}}
\end{minipage}
\caption{Maximum elevation angles, total daylight duration and available solar energy}
\label{fig:SE}
\end{figure*}

The closed-form optimal solutions of \textbf{P2(a)} and \textbf{P2(b)} are iteratively updated using the alternate optimization technique to solve the original problem \textbf{P2} as detailed in Algorithm \ref{Algo2}. It begins by setting tolerance $\delta$ and a feasible $H$ at the chosen instance $n$. Then, we solve \textbf{P2(a)} for a given $H[n]$ to obtain the optimal $V[n]$, which is then used to solve \textbf{P2(b)} to update $H[n]$. This process repeats itertaively until the convergence criterion is met.
The minimized $P_{\rm req}$ is deducted from the stored energy yielding a fixed transmission power $P_{\rm T}$ for the night time communication. Given a requisite $P_{\rm T}$, we can again use the optimal power allocation strategy to maximize the sum rate within the power budget as detailed in the solution of \textbf{P1(a)}.




%

\section{Numerical Results}
The numerical results are evaluated for each hour of SS and WS of the year 2023 at a location indicated by $\xi$ and $\chi$ in Table \ref{tab:Value}. Using this information, parameters $j_d$ and $\epsilon_s$ are computed using the Python's module Pysolar \cite{pysolar}\footnote{Julian date and solar elevation angles can alternately be calculated using the online calculators available at https://www.typecalendar.com/julian-date and https://keisan.casio.com/exec/system/1224682277, respectively}. The air pressure and air density at an altitude $h$ in \eqref{eq.rel_pressure} and \eqref{eq.rho}, respectively, are approximated with second degree polynomials using curve fitting for the desired altitude range as $ \bar{p}_h\left( h\right) = 60 h^2  -3276.7h  + 47022.8$ (Pa) and $\bar{\rho}_h = \left(0.95162 h^2 -52.29356 h + 753.39927\right){\rm x}10^{-3}
{\rm (kg/m^3)}$. 
These approximated $\bar{p}_h$ and normalized $\bar{\rho}_h$ are plotted against the actual $p_h$ and normalized $\rho_h$ for the desired altitude range as shown in Fig. \ref{fig:APT}. The fitted curves allow us to use the continuous functions in-place of the piece-wise functions in the optimization process. This figure also highlights the atmospheric characteristics of decreasing atmospheric pressure and air density while increasing temperatures with the increasing flight altitudes. These trends are helpful in deciding the optimal station-keeping flight parameters such as altitude and cruise speed with the minimal power consumption.  
We assume a station-keeping flight radius in range $ 2.5 {\rm km}\leq r \leq 4 {\rm km}$ which guarantees $99.00\%$ to $99.95\%$ availability over a certain coverage area. Moreover, we 
presume compound multi-junction Gallium Arsenide (GaAS) solar cells mounted with efficiencies of  $39.1\%$ under one sun illumination\cite{yamaguchi2020high}.  We consider  one center cell and $6$ edge cells \cite{Nokiahsieh2020uav} with uniformly distributed users at elevation angles defined by the boundaries $\psi_e =\pi/15$ and $\psi_c = \pi/6$ in the $100$km coverage radius. These choices are based on the required coverage area as well as the channel characteristics i.e., the LoS link is not available in the region beyond $\psi_e =\pi/15$ resulting in a Rayleigh channel instead of Ricean channel. Moreover, we assume the center cell beam with $\theta^b = \pi/2$ whereas a highly directional beam with $\theta^b = \pi/15$ for the surrounding edge cells. The adopted values of numerous simulation parameters are presented in Table \ref{tab:Value} unless specified otherwise.

 \begin{table}[t]
\caption{The Used Choices of the System Parameters}\label{tab:Value}
\centering
  \begin{tabular}{||c|c||c|c||}
     \hline
{$\xi$}&$2^0$ $14$' $2.04$"E& $M$,$r$ &$ 7$,$3$km\\
          \hline
{$\chi$}&$53^0$ $28$' $0.48$"N& $K_m$&$30-3000$ \\
          \hline
  SS: {$j_d$}& $2460117$ & $E_{\rm acc}$ & $2.3$kWh \\
          \hline
 WS: {$j_d$}&  $2460300$  &$S$ & $143 {\rm m}^2$ \\
          \hline
SS: $\alpha_{\rm ext}$ & $0.465$ & $W$ &$165 {\rm kg}$\\
     \hline
WS: $\alpha_{\rm ext}$ & $0.29$ &$C_{D_{0}}$&$0.015$\\
 \hline
$b$,$AR_w$&$35m,30$&$C_{\rm Lmax}$&$1.2$\\    
     \hline
$B$&$20$MHz&$e$&$0.6385$\\
     \hline
$T_n$&$870$&$\delta $, NF&$ 1e-4$, $5$dB\\
    \hline
   $k_B$,$ \Upsilon$&$1.3800e-23$,$10\%$ &$M_b,m_b$&$18,-2$dBi \cite{shibata2020system}\\ 
        \hline
   $f_c$&$2.1$GHz &$\sigma_f^2,\Upsilon $&$1,0.1$\\
          \hline
    $\eta_p,\eta_e$ & $0.85, 0.90$&$G_r$&$1$\\
           \hline
  $\eta_c , \eta_d$ &$0.93, 0.97$   & $\eta_s $&$0.29$\\   
  \hline
         \hline
  \end{tabular}
\end{table}

\begin{figure*}[!t]
\hspace{1.5cm}
 \begin{minipage}[b]{0.4\textwidth}
  \centering
 {{\includegraphics[width=\linewidth]{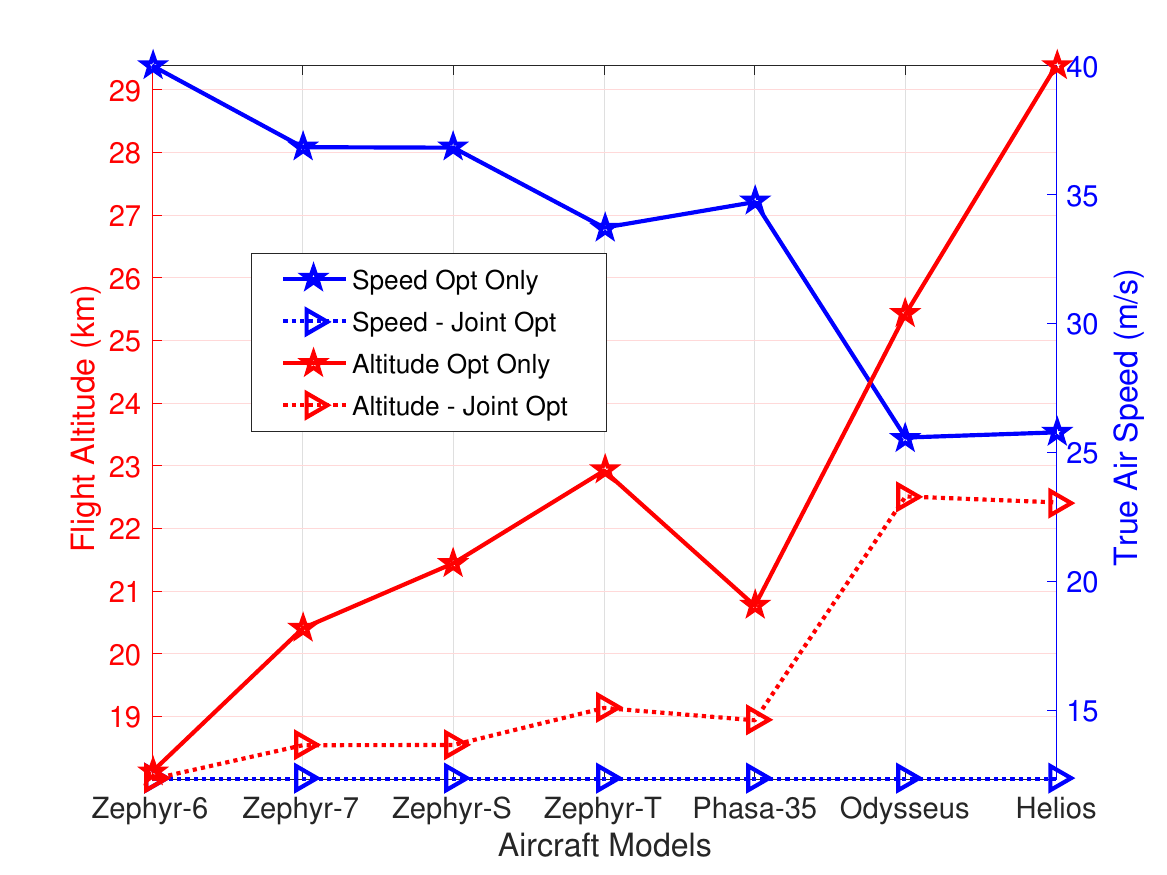}}}
\end{minipage}
\begin{minipage}[b]{0.4\textwidth}
  \centering
{\includegraphics[width=\linewidth]{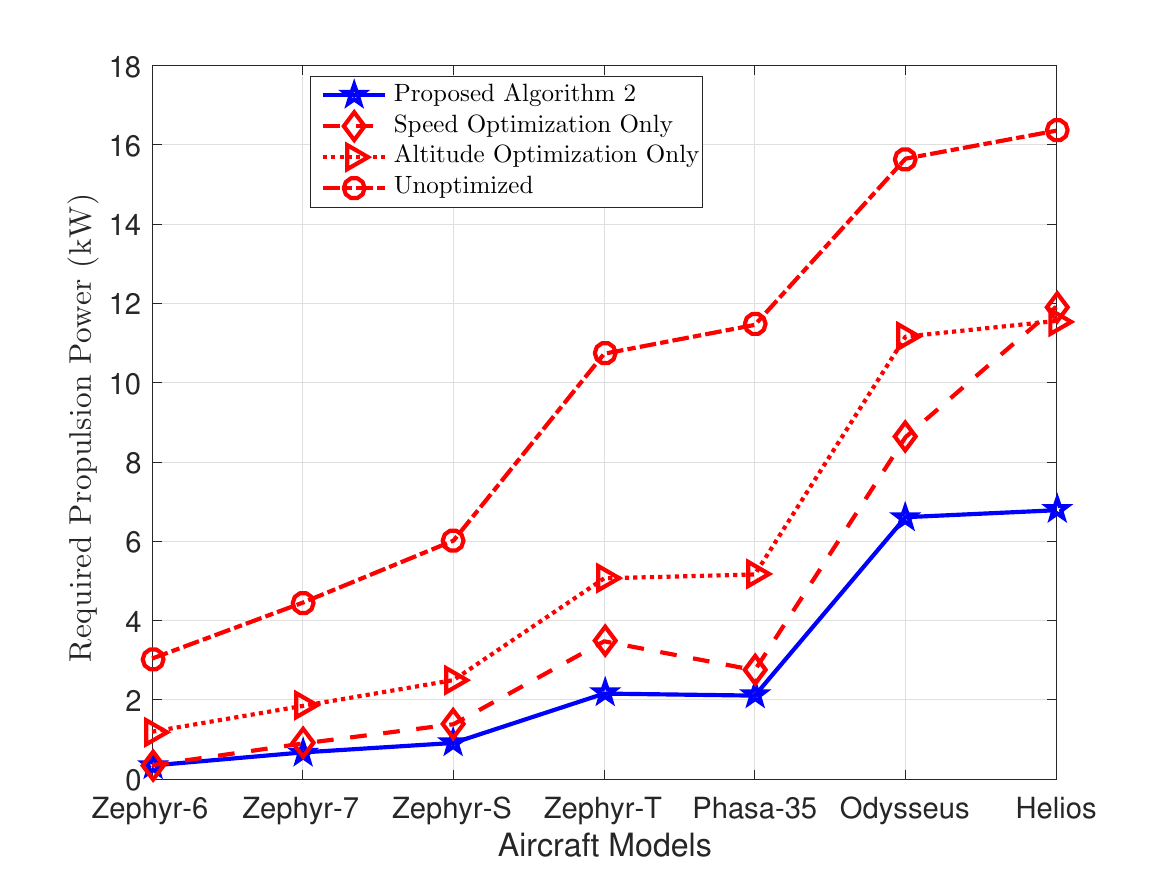}}
\end{minipage}
\caption{Altitude, Speed  and Propulsion Power Requirement for Different Aircraft Models}
\label{fig:Prop}
\end{figure*}
The maximum elevation angle on each day of the year is evaluated for four different locations with increasing latitude from the equator as shown in Fig. \ref{fig:SE}(a). The distant places from the equator depict lower elevation angles as compared to the closed ones. The unique trend followed by the $\epsilon_{\rm sm}$ in Malaysia and KAUST can be explained by knowing the maximum achievable elevation angle i.e., $90^o$. Similarly, Fig. \ref{fig:SE}(b) highlights the daylight duration on each day of the year. The maximum and minimum daylight durations of around $20$ and $4$ hours are observed in Alaska, respectively.  This emphasizes the significance of location parameters and daylight hours while designing the HAPS system. A self-sustaining HAPS should be able to survive with only $4$-hours of daylight if it has to serve in Alaska. On the other hand, the regions around equator enjoy a consistent daylight duration throughout the year. Eventually, the total solar energy per unit area is illustrated in Fig. \ref{fig:SE}(c), it is dependent on both $\epsilon_{\rm sm}$ and $\tau$. Clearly, the maximum solar energy is available for the regions close to equator; however, the solar energy varies dramatically for the farthest regions. A successful HAPS design should be able to perform its routine tasks with only $1$kWh/$m^2$ during the winters in Northern regions.

Next, we evaluate our proposed algorithm for the optimal flight speed and altitude to minimize propulsion power during the night for the case study of seven different candidate HAPS. Various companies like Boeing, Airbus, NASA, and BAE systems, etc., have designed and tested their prototypes for aerial communication from stratosphere. We have opted Zephyr-$6$, Zephyr-$7$, Zephyr-S, Zephyr-T, PHASA-$35$, Odysseus, and Helios aircraft models with wingspan  $[18.28; 22.5; 25; 33; 35; 74; 75.3]$ meters, total weight (platform and payload mass) $[32; 58; 75; 160; 165; 700;  720]$ kg, wing area $[38.9364; 56.25; 76; 133; 143; 175; 183.6] {\rm m}^2$, and maximum achievable altitude $[18.3; 21.6;23.2;24;21.336;25.5;29.5]$ km, respectively. Pertaining to the lack of information regarding the true airspeed, we assume $V_{\rm min}=12m/s$ and $V_{\rm max}=40m/s$ whereas $V_s$ varies as per the dimension of each aircraft.
Fig \ref{fig:Prop}a illustrates the optimal speed and altitude in three different scenarios i.e., speed optimization only at a given altitude ($H_o$) , altitude optimization only at a given airspeed ($V_o$) and our proposed algorithm for joint optimization of HAPS speed and altitude.
Interestingly, lighter aircrafts (Zephyr) would fly with higher airspeed whereas heavier aircrafts (Helios) would fly with lower airspeed under speed optimization scenario. However, the optimum result of joint optimization is lowest possible airspeed to maintain a SCF in diameter $4{\rm km}$ to reduce propulsion power.  On the other hand, the optimum flight altitudes of each aircraft are dependent on their maximum attainable altitude during the NTO, whereas they cruise at lowest possible altitude i.e., $18{\rm km}$ during the DTO to minimize large scale-fading and maximize the sum rate. On the other hand, Fig \ref{fig:Prop}b presents the propulsion power requirement of all aircrafts in individual optimization, joint optimization and the worst case scenario of unoptimized parameters. Evidently, the flight requires minimum propulsion power for all aircrafts with the joint optimization approach. However, in terms of one parameter optimization, speed optimization is superior to altitude optimization as it renders the lower propulsion power, as shown in Fig \ref{fig:Prop}b. The joint optimization render a percentage reduction of $88.6759\%,   84.8433\%,   84.8169\%,   79.8841\%,   81.6073\%$,   $57.7812\%,   58.5283\%$ in propulsion power requirement as compared to the one with unoptimized flight parameters in respective aircraft models.

\begin{figure*}[!t]
 \begin{minipage}[b]{0.333\textwidth}
  \centering
 {{\includegraphics[width=\linewidth]{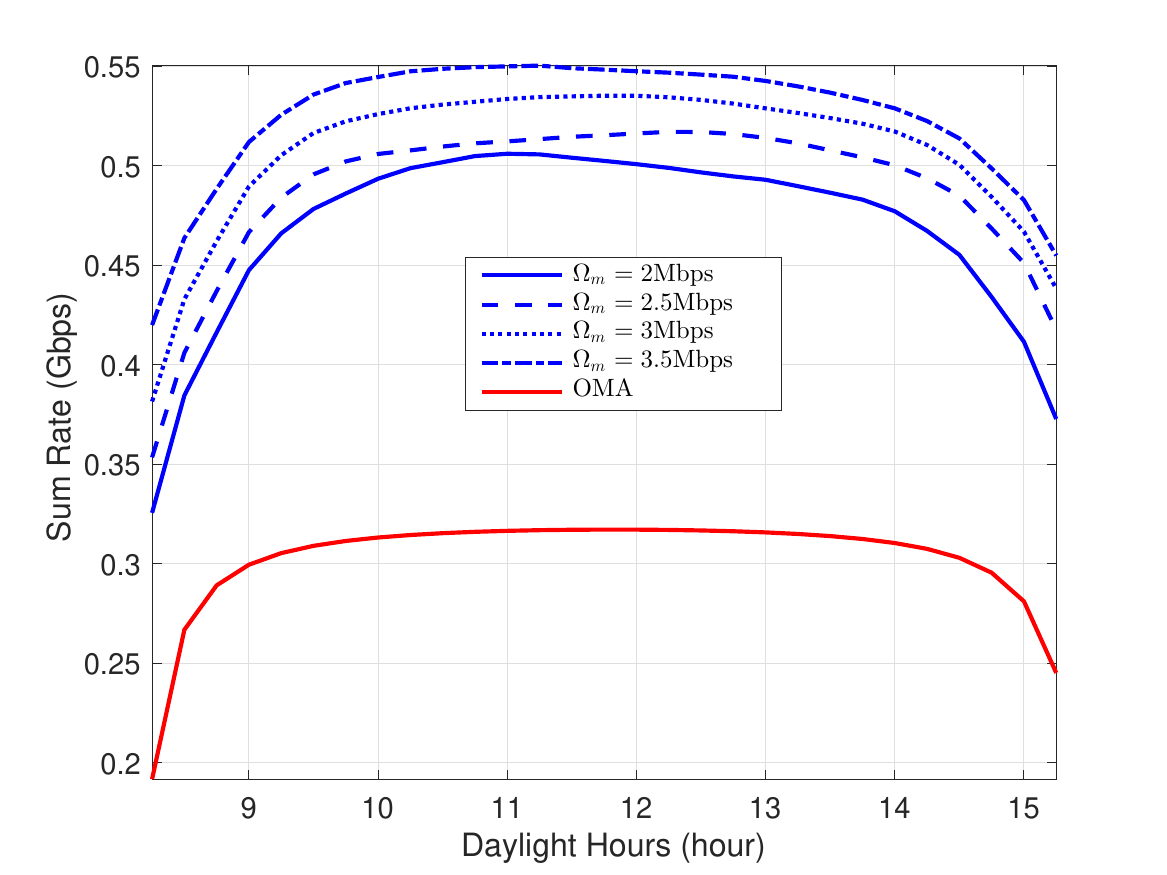}}}
\end{minipage}
\begin{minipage}[b]{0.333\textwidth}
  \centering
 {{\includegraphics[width=\linewidth]{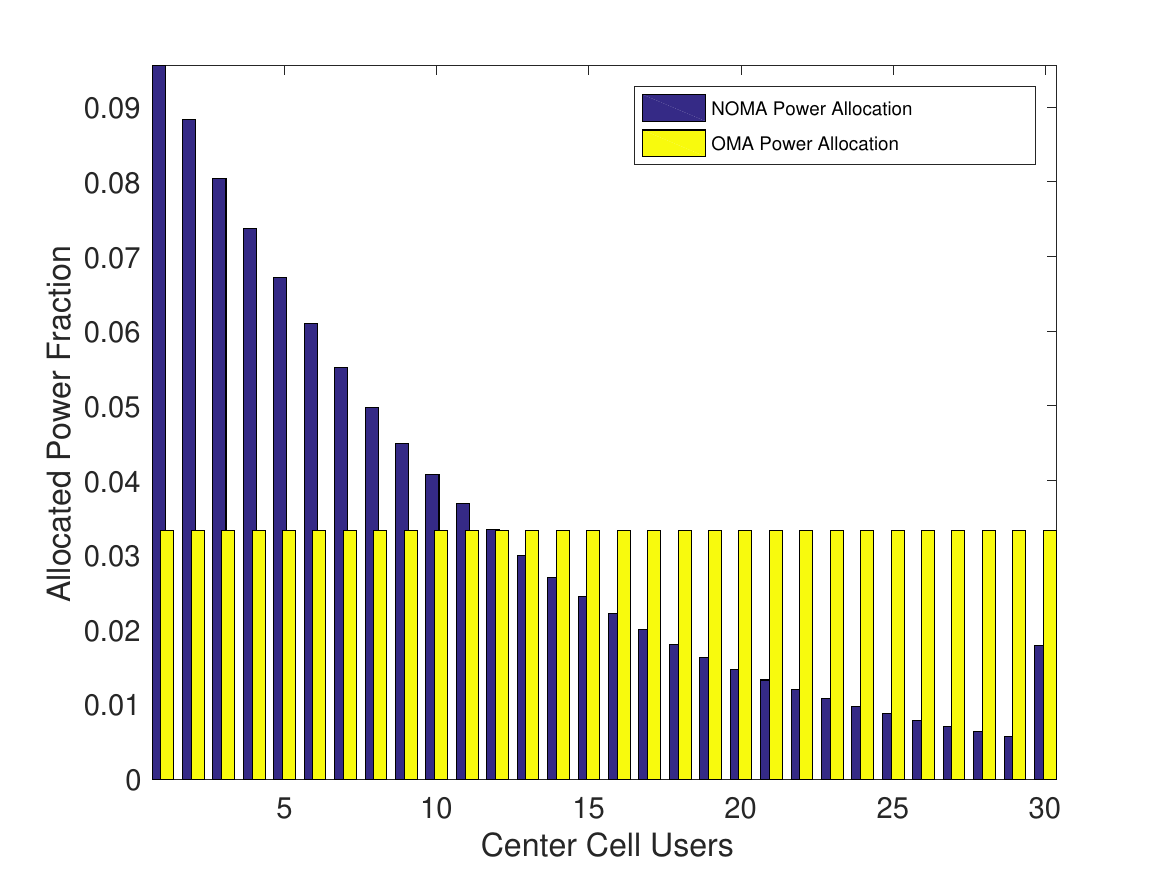}}}
\end{minipage}
\begin{minipage}[b]{0.333\textwidth}
  \centering
{\includegraphics[width=\linewidth]{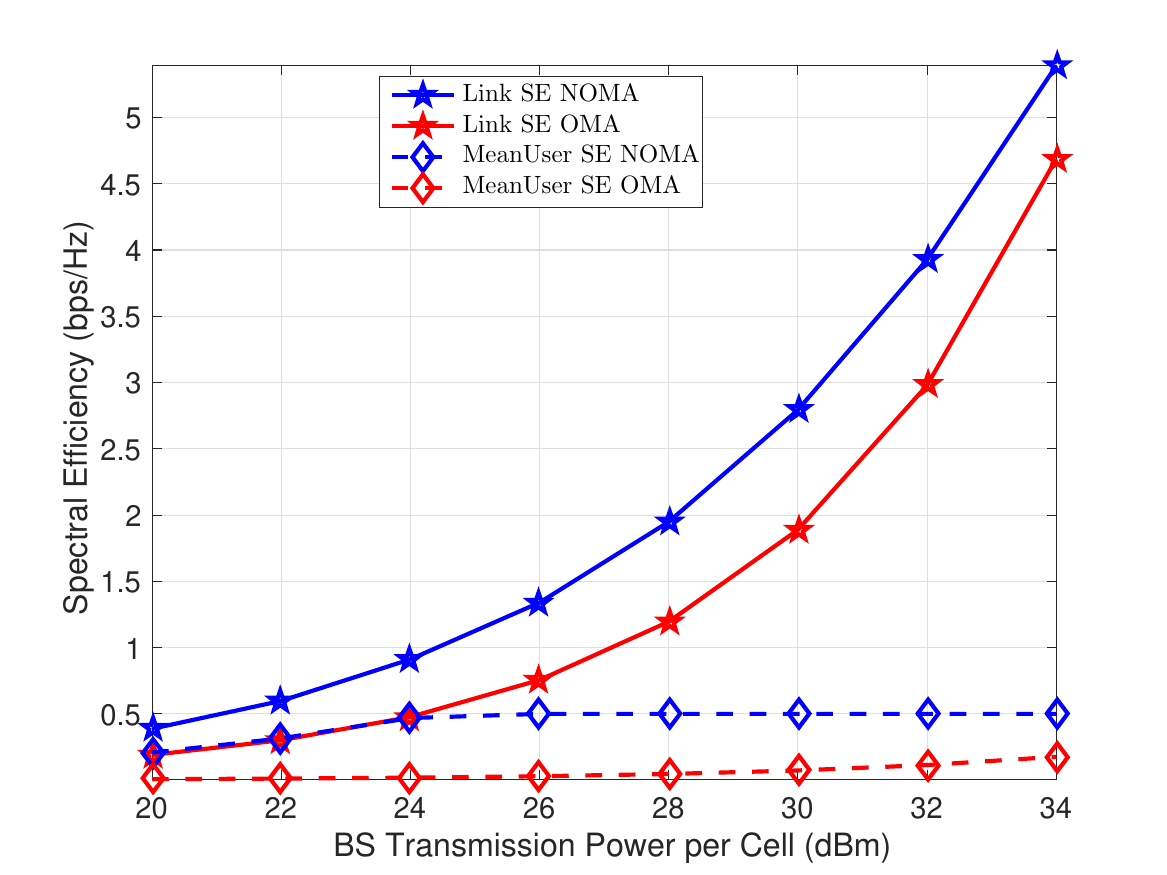}}
\end{minipage}
\caption{Average Sum Rate, Power Distribution and Spectral Efficiency}
\label{fig:SR}
\end{figure*}

In the next scenario, we present the average sum rate of $210$ users in the coverage area of $100$km radius with optimal power allocation in the daylight hours. We assume PHASA-35 flying over Manchester in the worst case i.e.,  winter solstice with merely $7$hours of daylight.  We propose utilizing all the excessive power for wireless transmission after meeting the necessary propulsion, accessory expenses and storing the requisite energy for NTO. We break the daylight hours in equal time interval of $15$ minutes as the solar elevation angle is almost constant in this duration. We use the available solar energy values over Manchester and optimal speed, altitude and propulsion power of PHASA-35 from Fig. 6 and 7, respectively 
The average sum rate with the optimal power allocation in NOMA clearly outperforms the OMA counterpart with equal power distribution while ensuring four different QoS rate thresholds as illustrated in Fig. \ref{fig:SR}a. Evidently, users can achieve a higher sum rate around noon as compared to the rest of the day pertaining to the higher $P_m$ available at that time. Interestingly, the optimized system renders $ 69.8340\%$, $84.3814\%$, $99.0371\%$, and  $119.0802\%$ relative gain in the average sum-rates over the unoptimized system with $\Omega = 2, 2.5, 3, {\rm or} \; 3.5$Mbps  QoS thresholds, respectively.

The bar chart in Fig. \ref{fig:SR}b demonstrates the power distribution amongst users in the center cell with NOMA as opposed to the uniform division with OMA. The users with weaker channel strength (edge cell users) are allocated higher fraction of available power budget as compared to the users with stronger channel conditions (center cell users) for a fair distribution enabling every user in the cell to meet QoS threshold. Once all users are able to meet the target rate, the excessive power is allocated to the strongest user i.e., $u_{30}$, which maximizes the overall sum rate as discussed in Theorem 1. On the other hand, for higher QoS threshold only few users with stronger channel conditions can achieve the targets while the weak users experience outage.

Finally, the link spectral efficiency (LSE) and mean user spectral efficiency (MUSE) are illustrated for the similar case of NOMA and OMA (power and bandwidth allocation) for users in the service area of PHASA-35 flying over Manchester on WS \footnote{The choice of Manchester is to highlight the sustenance and year long operation of HAPS at such high latitude. From the given set, Malaysia and KASUT already enjoy abundant solar irradiance whereas Alaska with limited number of inhabitants cannot sustain year-long operation owing to the very few daylight hours during winters}. We define LSE as the ratio between sum rate of all users and total bandwidth whereas MUSE is the mean of the ratio between the average user rate and bandwidth of all users in the center cell. We can observe upto $50\%$ and $26.7\%$ percentage improvement in the LSE and MUSE, respectively, by using NOMA over OMA.

\section{Conclusion}
A self-sustaining long-endurance HAPS has been proposed to serve a large coverage area while cruising in a stratospheric station-keeping flight.
We have presented an interdisciplinary approach to tackle the challenges in solar energy harvesting, aerodynamics and communications for efficient power budgeting.  We further proposed downlink NOMA strategy to serve  multi-cell users from the array panels (mounted underneath the flying base station in HAPS) for efficient resource management. 
The propagation model and link budgeting involved both the small and large scale fading for accurate performance analysis. 
We formulated different optimization problems for the day and night operations with the objectives to maximize sum-rate and minimize power requirements, respectively.  We furnished Algorithm I and II to describe the optimization procedure based on the closed form solutions of the sub-problems. Our findings of the numerical results emphasized the 
joint design of flight and transmission parameters to maximize sum rate and minimize propulsion power requirement. We have achieved upto $88.67\%$ percentage reduction in the required propulsion power of Zephyr-$6$. Moreover, we have obtained upto $119.08\%$ and $50\%$ percentage improvement in the average sum rate and spectral efficiency, respectively, using NOMA over OMA counterpart while satisfying QoS thresholds of the users. Thus, stratospheric aerial communication platforms can emerge as the promising candidates for global coverage without requiring any special UE.

\appendices
\section{Derivation of $H_m$ and $V_m$}
The gradient of propulsion power $P_{\rm pro}$ \eqref{eq.Prop} with respect to altitude $H$ is given by
\begin{equation} \label{eq.partialH}
\frac{\partial P_{\rm req} }{\partial H}=\frac{1}{\eta_p \eta_e} \left( \frac{1}{2}  \frac{\partial \rho_h }{\partial H}  V^3  S C_{D_0} - \epsilon \frac{2W^2}{\rho_h^2 S V} \frac{\partial \rho_h }{\partial H}\right)
\end{equation}
Equating \eqref{eq.partialH} to zero yields the following
\begin{equation} 
 \rho_h \left(H^*\right) =  \frac{2W}{ S  V^2}  \sqrt{\frac{\epsilon}{C_{D_0}}}
\end{equation}
We can then obtain the optimum $H_m$ \eqref{eq.Hm} using the piecewise function $\rho_h$  for Stratosphere-I and Stratosphere-II.
Similarly, the gradient of propulsion power with respect to airspeed is given as
\begin{equation} \label{eq.partialV}
\frac{\partial P_{\rm req} }{\partial V}= \frac{1}{\eta_p \eta_e} \left( \frac{3}{2}  \rho_h  V^2  S C_{D_0} - \epsilon \frac{2W^2}{\rho_h S V^2} \right)
\end{equation}
The stationary point obtained by setting \eqref{eq.partialV} to zero renders the optimal $V_m$ in \eqref{eq.Vm}.

 \bibliographystyle{IEEEtran}
\bibliography{IEEEabrv,refs}

\end{document}